\documentclass[%
 reprint,
superscriptaddress,
nobibnotes,
 amsmath,amssymb,
prx,
]{revtex4-1}

\usepackage{graphicx}% Include figure files
\usepackage{dcolumn}% Align table columns on decimal point
\usepackage{bm}% bold math
\usepackage{siunitx}
\usepackage{hyperref}% add hypertext capabilities
\begin{document}

\title{Spin Spectroscopy of a Hybrid Superconducting Nanowire \\ Using Side-Coupled Quantum Dots}% Force line breaks with \\

\author{Alisa Danilenko}
\affiliation{Center for Quantum Devices, Niels Bohr Institute, University of Copenhagen, 2100 Copenhagen, Denmark}

\author{Andreas P\"oschl}
\affiliation{Center for Quantum Devices, Niels Bohr Institute, University of Copenhagen, 2100 Copenhagen, Denmark}

\author{Deividas Sabonis}
\affiliation{Center for Quantum Devices, Niels Bohr Institute, University of Copenhagen, 2100 Copenhagen, Denmark}

\author{Vasileios Vlachodimitropoulos}
\affiliation{Center for Quantum Devices, Niels Bohr Institute, University of Copenhagen, 2100 Copenhagen, Denmark}%

\author{Candice Thomas}
\affiliation{Department of Physics and Astronomy, and Birck Nanotechnology Center, Purdue University, West Lafayette, Indiana 47907 USA}

\author{Michael J. Manfra}
\affiliation{Department of Physics and Astronomy, and Birck Nanotechnology Center, Purdue University, West Lafayette, Indiana 47907 USA}
\affiliation{School of Materials Engineering, and School of Electrical and Computer Engineering, Purdue University, West Lafayette, Indiana 47907 USA}

\author{Charles M. Marcus}
\affiliation{Center for Quantum Devices, Niels Bohr Institute, University of Copenhagen, 2100 Copenhagen, Denmark}

\date{\today}% It is always \today, today,

\begin{abstract}
We investigate superconducting hybrid nanowires defined by patterned gates on a two-dimensional heterostructure of InAs and Al, with lateral quantum dots operating as single-level spectrometers along the side of the nanowire. Applying magnetic field along the wire axis spin splits dot levels, providing spin-resolved spectroscopy. We investigate spin and charge polarization of subgap states in the nanowire and their evolution with magnetic field and gate voltage.

%\begin{description}

\end{abstract}

\maketitle

%\tableofcontents

\section{Introduction}
\begin{figure*}[t]
\includegraphics[width=\textwidth]{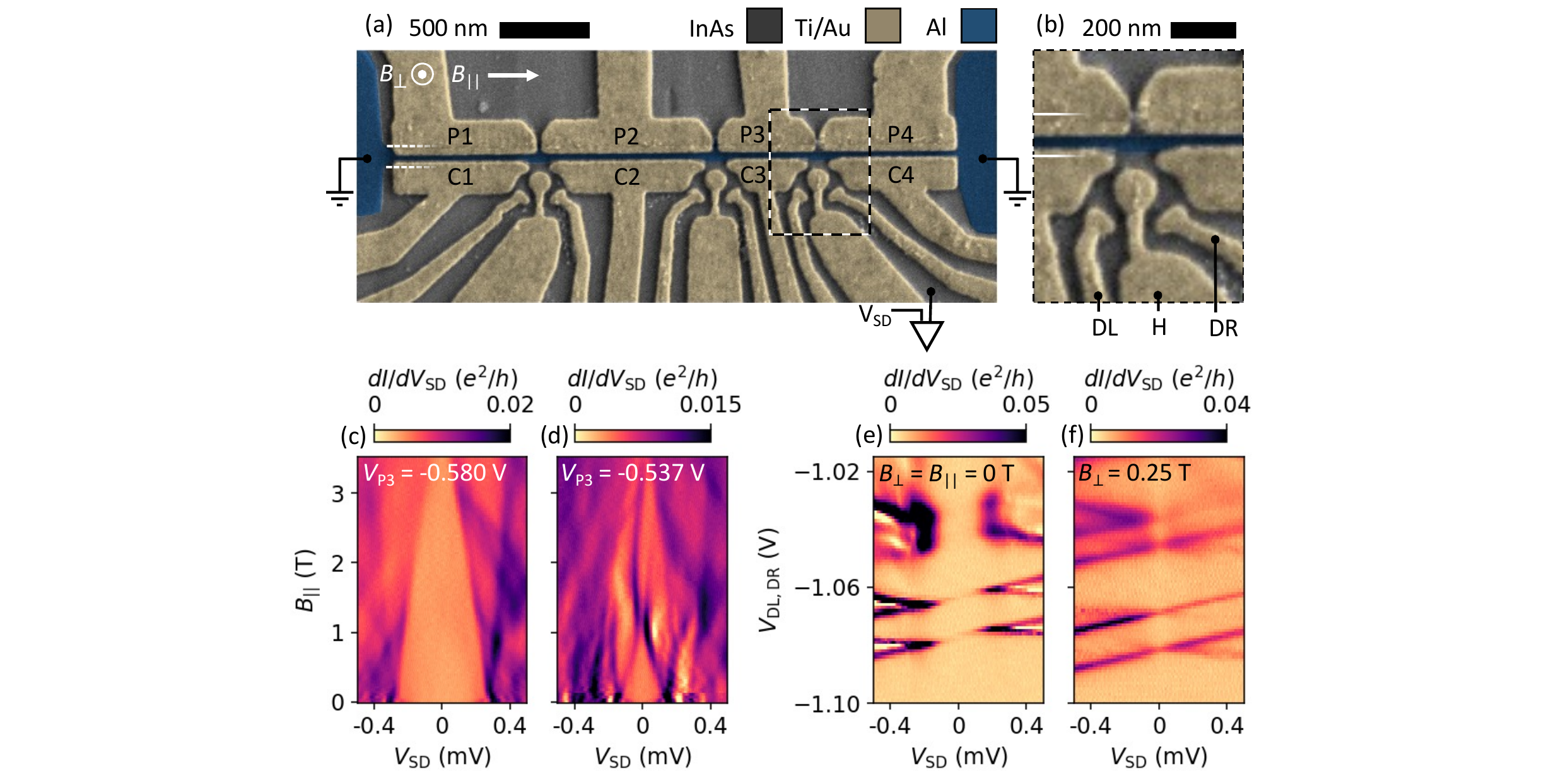}
\caption{\label{fig:device} (a) False-color electron micrograph of a device identical in design to device 1,
formed around an etched strip of epitaxial Al on an InAs quantum well. Gates P and C deplete  carriers in the 2DEG on either side of the Al, forming a proximitized NW. Gates can additionally be used to tune the chemical potential in the NW, segment by segment. Gates labelled C also form tunnel barriers to normal leads at three locations along the NW, where voltage amplifiers are connected. Relevant field directions are marked in white; parallel to the NW ($B_{||}$) and perpendicular to the plane of the 2DEG ($B_\perp$). (b) A magnified view of the third probe location (dashed box in (a)). Gates DL and DR define a QD at the probe location, and are used to tune the coupling between the QD and the normal lead. Gate H allows for additional control over the QD. (c, d) Tunneling spectroscopy measured at the third probe location as a function of $B_{||}$, for two different values of $V_{\mathrm{P3}}$. (c) Closing of the superconducting gap as a function of magnetic field is observed without subgap features. (d) A bound state in the gap splits and an anticrossing is observed. During these measurements the voltages on the QD-related gates (DL, H, DR) are set to 0~V, so there is no QD in the probe location, just a tunnel barrier formed by the gates C3 and C4. (e, f) Coulomb blockade measurements of the QD formed by energizing DL and DR, at zero and finite $B_\perp$ respectively.  }
\end{figure*} 

When indium arsenide (InAs), a semiconductor, is coupled to aluminum (Al), a superconductor, the two materials inherit properties from each other, effectively creating a new material system \cite{kroger_superconductor-semiconductor_1989, krogstrup_epitaxy_2015, shabani_two-dimensional_2016}. The proximity effect induces effective pairing in the InAs \cite{takayanagi_superconducting_1985} via Andreev reflection from the superconductor \cite{beenakker_quantum_1992, schapers_proximity_2001}, opening a gap in the spectrum of the otherwise semiconducting system \cite{chrestin_soc_1997}. A large $g$-factor and spin orbit coupling in the hybrid system are inherited from the InAs \cite{fasth_direct_2007, shabani_two-dimensional_2016, oconnell_yuan_epitaxial_2021}. 

One platform in which structures of this kind can be realized, and complex device geometries can be fabricated in a scalable manner, is a two-dimensional electron gas (2DEG) proximitized by a superconducting layer \cite{kjaergaard_transparent_2017, shabani_two-dimensional_2016}. If these hybrid systems are restricted to one dimension by gating, they become a hunting ground for a range of quantum states including Yu-Shiba-Rusinov, Andreev bound states (ABSs), and Majorana bound states (MBSs) \cite{balatsky_impurity-induced_2006, whiticar_parity_2021, henri_lead, whiticar_coherent_2020, nichele_scaling_2017, ofarrell_hybridization_2018,chang_tunneling_2013, jellinggaard_ysr_2016, kurtossy_andreev_2021}. These quantum states possess properties such as spin and electron-hole polarization, which respond to experimental parameters. There have been several proposals for the use of quantum dots (QDs) to probe NW state properties to elucidate their parity, spin texture, and localization \cite{clarke_experimentally_2017, prada_measuring_2017, penaranda_quantifying_2018}. Corresponding experimental efforts have already enabled investigation of the spatial extent of bound states using strongly coupled QDs which hybridize with the bound states \cite{deng_nonlocality_2018, PoschlPC2022}.

Previous experiments used a weakly coupled QD to read out the size of the superconducting gap in a proximitized region \cite{junger_dot_2019}, and to probe the above-gap resonances in the density of states (DOS) of a similarly proximitized system \cite{thomas2021} as well as transport through subgap resonances \cite{gramich_subgap_2016}. The use of QDs as spin filters has been exploited in the context of spin qubits \cite{hanson2003}, and their charge filtering properties have been utilized to probe the quasiparticle charge and energy relaxation in hybrid structures \cite{wang_nonlocal_2022}. However, QDs have not yet been used to address the spin and charge degrees of freedom of discrete subgap states in hybrid NWs. A similar study paralleling ours using InSb nanowires is reported in van Driel \textit{et al}. \cite{vanDriel}. 

In this Article, we introduce a device geometry based on an InAs/Al heterostructure that allows for laterally defined QDs that are side-coupled to a quasi-1D hybrid NW at multiple probe locations. The QDs are defined by electrostatic gating. The gate configuration allows the QDs to be weakly coupled to the NW, acting as a non-invasive probe of the local DOS at various points along the wire. When a magnetic field is applied to the system, the QD energy levels are split, providing spin-selective probes. We take advantage of this, using the QD levels to measure the spin splitting of the superconducting gap \cite{meservey1994}, extracting a $g$-factor of $\sim 1.7$, consistent with QPC-based tunneling spectroscopy measurements of the system. We further investigate a bound state in the NW, using sequential tunneling spectroscopy through spin-split QD levels to extract the magnitude and relative sign of the $g$-factor of the state. Finally, we present quantitative measures of spin and charge polarization of the current into the bound state measured via the QD spectrometer, and examine these quantities as a function of parallel magnetic field and chemical potential. 

If the gates forming the QDs are not energized our gate geometry allows for regular tunnelling spectroscopy through a quantum point contact (QPC)-like potential barrier, allowing separate  measurements of the subgap energy without spin filtering. These results are found to be consistent with QD spectroscopy. Measurements from two lithographically similar devices are presented. 

The Article is structured as follows: First, the device design is introduced (\ref{design}). Next, the use of QD levels as spectrometers is considered at low magnetic field (\ref{lowfield}) then at high magnetic field (\ref{highfield}). Finally, the spin and particle-hole polarization of tunneling current through the QD-bound state system is investigated (\ref{spinres}), and the results are discussed in the context of the current literature (\ref{discuss}).

\begin{figure*}[t]
\includegraphics[width=\textwidth]{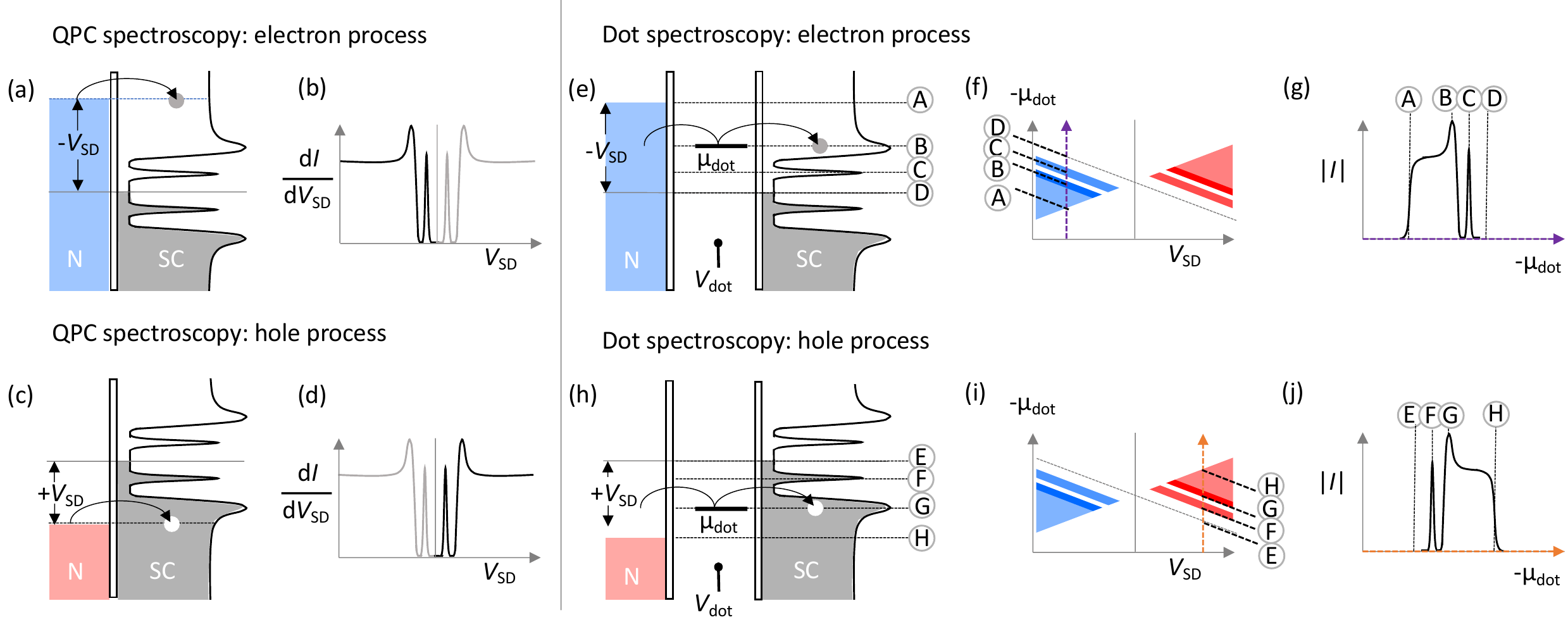}
\caption{\label{fig:zerotheory}Schematic illustration of the use of a single level of QD as a spectrometer to measure the DOS of the proximitized NW for the case of one subgap state at zero magnetic field. (a, c) Sketches of the electrochemical potentials involved in conventional tunneling spectroscopy at negative and positive voltage bias. A bias voltage is applied to the normal lead and differential conductance is measured through a tunnel barrier between the lead and the NW, giving a signal that is  proportional to the DOS as a function of the bias voltage (b, d). (e, h) A single QD level is included in the diagram, with fixed negative and positive voltage bias on the lead. The energy of the QD level can be tuned using a gate voltage. The points labelled A, B, C, and D in (e) and E, F, H, and H in (h) represent positions where the QD level is on resonance with the normal lead, the superconducting coherence peak, the ABS in the NW, and zero bias.  Schematic Coulomb diamonds expected  for these configurations are shown in (f) and (i). Slices through the Coulomb diamond are sketched at finite negative (g) and finite positive (j) bias, with special points marked to illustrate the action of the spectrometer.}
\end{figure*} 

\section{Device design}\label{design}

Results from two devices are reported. A scanning electron micrograph of a device identical in design to device 1 is shown in Fig.~\ref{fig:device}; device 2 is structurally similar. The device is fabricated on an InGaAs/InAs/InAlAs heterostructure covered with {\it in-situ} grown $5$ nm of epitaxially matched Al. Following a mesa etch, an additional wet etch is used to define an Al stripe of  $\sim~100$ nm in width and $\sim~4$  $\mu$m in length, which extends into large ground planes at both ends. A layer of $15$~nm HfO$_\mathrm{x}$ is deposited globally and functions as gate dielectric. Ti/Au gates are then evaporated in two lithographic steps, one thin layer for fine features and a thick outer layer that crawls over the mesa and makes contact with the thin layer. Gates labelled P and C are used to deplete the carriers in the 2DEG self-aligned with the strip of Al, so that a quasi-1D proximitized channel is defined. The gates separate the NW into segments so that different segments can be tuned to have different chemical potentials by changing the applied gate voltage, allowing some control over the spatial distribution of bound states in the system. Additionally, the gates labelled C are used to define tunnel barriers at three locations along the NW. The planes of 2DEG, separated by the depletion of the C gates from the NW, are used as normal conducting leads.

The results highlighted in this Article will focus on measurements in the section of the device shown by the dashed box in Fig.~\ref{fig:device}(a). A close up image of this region, with the QD coupled to the NW from the side, is shown in Fig.~\ref{fig:device}(b). By depleting with the gates C3 and C4 (leaving gates DL, H, and DR at $0$ V), a QPC-like potential barrier is formed, through which differential conductance can be measured using standard lock-in techniques. Differential conductance measurements as a function of magnetic field parallel to the NW ($B_{||}$) are shown in Figs.~\ref{fig:device}(c, d) for two values of $V_{\mathrm{P3}}$. With $V_{\mathrm{P3}}=-0.580$~V the induced superconducting gap is seen closing in field with no subgap resonances, while at a slightly less negative gate voltage of $-0.537$~V there is a subgap state that splits in field and undergoes an anticrossing at $1$~T. Zero-bias gate-gate maps were taken at finite field to determine how strongly this state couples to different gates, indicating that the subgap state is localized under gate P3.

When gates DL and DR are energized with negative voltages, electron density is confined, forming a QD. The voltage configuration of these gates can be used to tune the coupling between the QD and the normal lead. The C gates can be further adjusted to tune the coupling of the QD to the NW. The H gate, with a circular part situated directly on top of the region where we expect the QD to form, can be used for further control over the QD. Using a combination of these gates, it is possible to tune over a range of coupling strengths, though towards the strong coupling limit this effect is somewhat nonmonotonic. Differential conductance measurements with the QD formed in the probe location are shown in Fig.~\ref{fig:device} for zero applied field (e) and $B_\perp=0.25$ T (f), with the latter field large enough to drive the Al of the NW normal. 

\section{Spectroscopy at low magnetic field}\label{lowfield}

To use a QD as a spectrometer to probe the DOS in the NW, it is necessary to operate in a strongly decoupled regime, where both the coupling between the QD and the superconductor ($\Gamma_S$) and between the normal lead and the QD ($\Gamma_N$) are much less than $k_B T$, such that inelastic co-tunneling processes are suppressed and sequential electron tunnelling dominates \cite{junger_dot_2019, gramich_andreev_2017, recher2000}. In this case, and provided $\Gamma_S<\Gamma_N$, the sequential dc current flowing into the NW is proportional to the DOS at an energy $E$ selected by the chemical potential of a single dot level \cite{junger_dot_2019}. The energy window for the measurement is selected by the dc source-drain bias voltage, $V_{\mathrm{SD}}$, and the level spacing of the QD must be larger than the selected energy window for spectroscopy to be performed. The concept is illustrated in Fig.~\ref{fig:zerotheory} for the case of a single subgap state. Panels (a) and (c) show sketches of the electrochemical potentials involved in standard spectroscopy through a tunnel barrier, for negative and positive voltage bias. In this case, a measurement of the DOS can be performed by varying the bias voltage applied to the normal lead, and recording the differential conductance. Figures~\ref{fig:zerotheory}(b,~d) show an example sketch of the differential conductance that would be measured in such a setup, for the case of one subgap state being present in the NW at finite energy. In contrast to this method, when using the QD as a spectrometer the bias on the normal lead can be kept at a constant value [Figs.~\ref{fig:zerotheory}(e, h)] and a single level of the QD can instead be swept by adjusting a gate, preferably one that dominantly tunes the QD chemical potential. A sketch of the current that would result, again for the case of one subgap state below a superconducting gap, is shown as a function of the potential of the QD level and the bias on the normal lead [Figs.~\ref{fig:zerotheory}(f, i)]. The role of the bias on the normal lead is to determine the energy window for the spectroscopy measurement. Panel (g) shows a sketch of the current along the fixed bias [purple arrow in panel (f)]. Points of interest are denoted A, B, C, and D. These correspond to the QD energy level being resonant with the source-drain bias, $V_\mathrm{SD}$, the superconducting coherence peak, a subgap state, and zero energy respectively, going from high to low $\mu_\mathrm{dot}$. Before reaching point A, the QD energy level is above the energy of the normal lead, so there is no tunnelling through the system and the measured current is zero. At point A, the QD level is resonant with the lead. Tunnelling becomes possible and a finite current proportional to the above-gap DOS switches on. At point B the QD level is resonant with the coherence peak of the induced superconducting gap, so a corresponding peak is observed in the measured current, and similarly at point C the peak corresponding to the finite bias state in the NW is observed. At point D, the QD level moves below the Fermi energy, so the measured current of electrons into the device vanishes once again. To measure the other half of the DOS, one needs to fix the bias at an equal magnitude but opposite sign [Fig. \ref{fig:zerotheory} (h)]. In this case, the current will flow in the opposite direction, which can be considered as an electron current with opposite sign, or a hole current into the device. The corresponding current along the fixed bias line denoted by the orange arrow in (i) is sketched in (j). 

\begin{figure}
\includegraphics[width=0.48\textwidth]{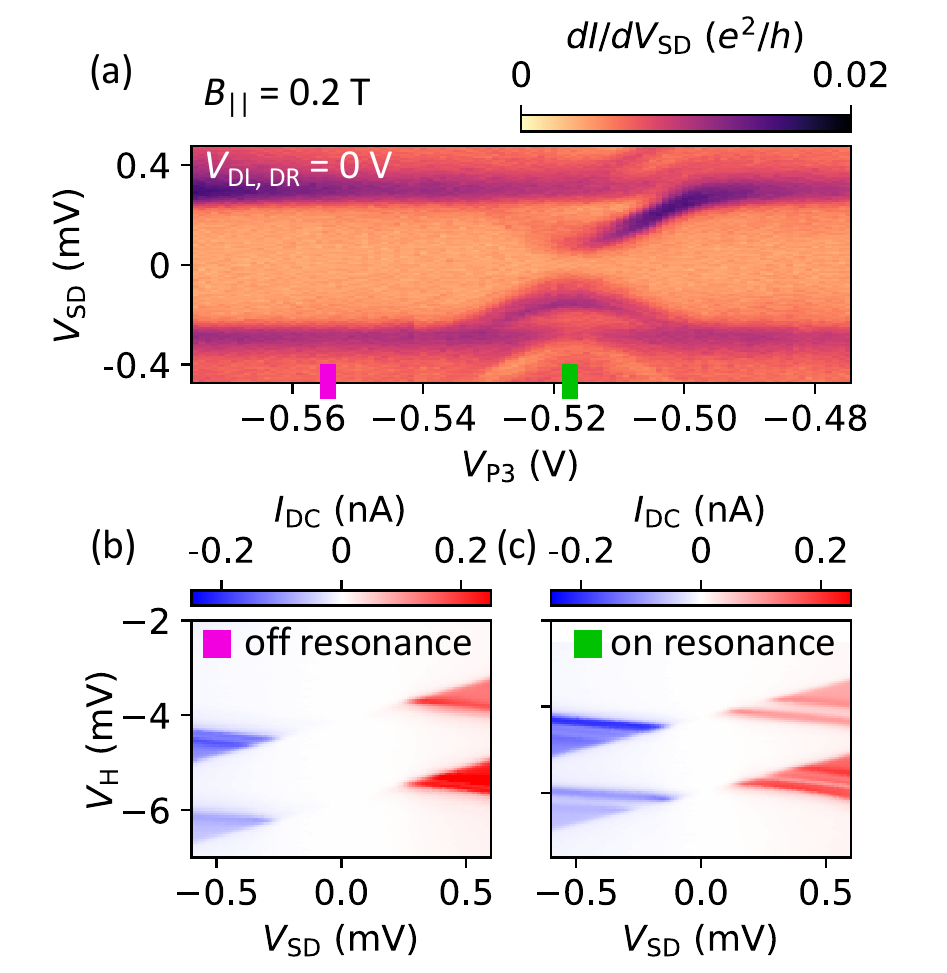}
\caption{\label{fig:datazerofield}(a) Tunneling spectroscopy at $B_{||} = 0.2$ T as a function of the NW plunger gate voltage $V_{\mathrm{P3}}$. A subgap state can be seen around $-0.52$ V. (b) Current measured through the weakly coupled QD when the gate voltage $V_{\mathrm{P3}}$ is set far away from the subgap resonance, indicated by a magenta square. (c) Current through the same two QD levels with $V_{\mathrm{P3}}$ set to the location of the subgap state energy minimum, indicated by a green square.}
\end{figure}

\begin{figure}
\includegraphics[width=0.48\textwidth]{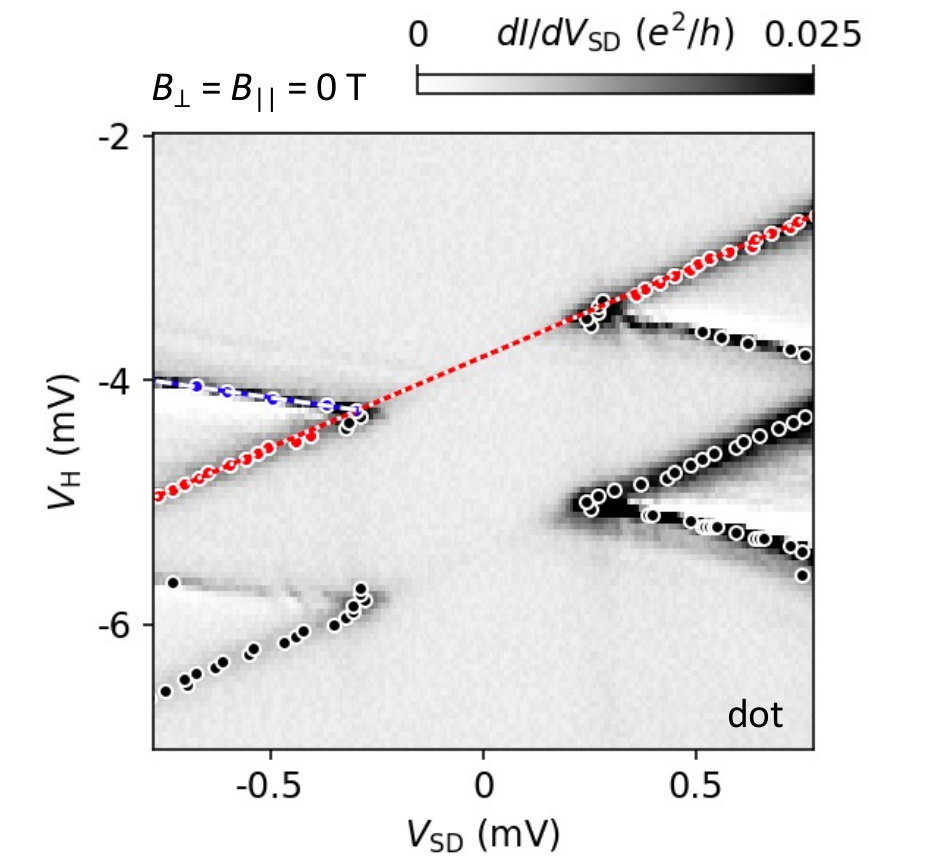}
\caption{\label{fig:leverarm} Method of extracting gradients to find the gate lever arm, shown for the two Coulomb resonances used for QD spectroscopy. Scatter points correspond to peaks at Coulomb diamond edges, found using the \texttt{scipy} {\it find peaks} function. Linear fits to points are used to find slopes $m_1$ (red points) and $m_2$ (blue points). Black points not used.}
\end{figure}

\begin{figure}
\includegraphics[width=0.48\textwidth]{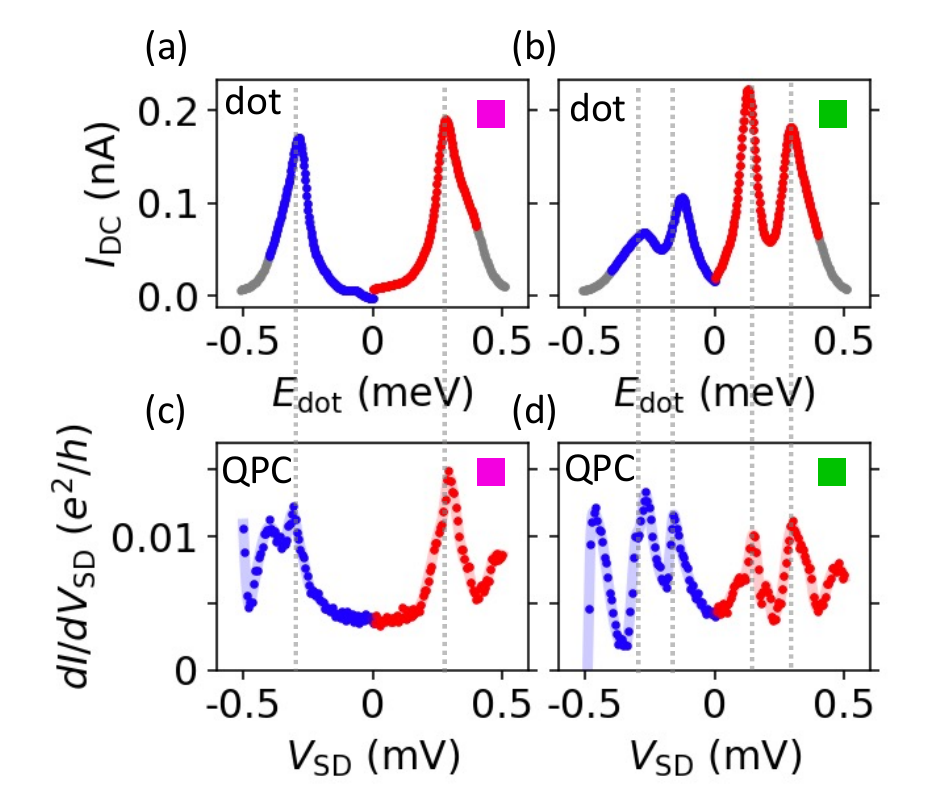}
\caption{\label{fig:1dcuts0field} Line cuts through a Coulomb diamond at $V_\mathrm{SD}=-0.4$~mV (blue) and $V_\mathrm{SD}=+0.4$~mV (red), showing the dc current, proportional to the density of states in the NW, as a function of the energy $E_{\mathrm{dot}}$. The $x$-axis was converted from $V_\mathrm{H}$ to $E_{\mathrm{dot}}$ using the gate lever arm $\alpha$. Shown at zero magnetic field for (a) an empty subgap and (b) with one ABS in the NW. (c) and (d) also show spectroscopy of the NW DOS in an empty-subgap region and for one ABS, but measured via tunnelling spectroscopy, by varying $V_{\mathrm{SD}}$ and measuring differential conductance. For these differential conductance measurements, the gates which form the QD are set to 0 V, so there are no QD levels to consider in the measurement.   }
\end{figure}

We observe a small splitting of all features in tunneling spectroscopy in a window of $\sim50$~mT around zero $B_{||}$. This splitting can be seen close to zero field in Figs.~\ref{fig:device}(c, d).  The cause of this is not clear, but we suspect it may be due to spin-orbit effects in the normal 2DEG leads. Because of this, for clarity, we demonstrate the low field action of the QD spectrometer at a small parallel magnetic field of $0.2$ T (away from the split features) instead of at zero field. Although there is some Zeeman splitting at this field value ($E_{Z, \mathrm{dot}}\sim 100$~$\mu$eV), the splitting of the QD levels is still small compared to the width of the DOS features measured, and the splitting of the superconducting gap and any subgap features is almost negligible ($E_{Z, \mathrm{gap}}\sim 20$~$\mu$eV, $E_{Z, \mathrm{ABS}}\sim 26$~$\mu$eV). Using the QPC with the D and H gates set to $0$ V, tunnelling spectroscopy on Probe 3 is measured [Fig.~\ref{fig:datazerofield} (a)] while sweeping the gate voltage $V_{\mathrm{P3}}$, which changes the density in the NW segment underneath it. A local state in the NW can be seen coming out of the continuum and into the gap, with a minimum at around $-0.52$ V. At $V_{\mathrm{P3}}$ values above and below the state resonance, there is a hard induced superconducting gap without subgap states. To demonstrate the action of the QD spectrometer at low magnetic field, a weakly coupled QD is formed by depleting with the D and H gates. We show Coulomb diamonds measured at two values of $V_{\mathrm{P3}}$: one at which there is no state inside the induced superconducting gap ($-0.53$ V, shown in Fig.~\ref{fig:datazerofield}(b)) and one at which the state in the NW reaches an energy minimum ($-0.5$~V, shown in Fig.~\ref{fig:datazerofield}(c)). These gate voltages are marked with magenta and green boxes, corresponding to the colored markers in Fig.~\ref{fig:datazerofield}(a). The $V_{\mathrm{P3}}$ values in Figs.~\ref{fig:datazerofield}(b, c) are not in one-to-one correspondence with the green and magenta box markers in Fig.~\ref{fig:datazerofield}(a) because the gates used to confine the QD have some capacitive coupling to the state in the NW, so turning on the QD spectrometer shifts the state slightly in $V_{\mathrm{P3}}$ space. For both cases, the Coulomb diamond structure looks as expected from our description of the sequential tunnelling path through the lead-QD-NW system. The absolute value of the dc current is non-zero only when an energy level of the QD falls in between the applied bias voltage $V_{\mathrm{SD}}$ and the Fermi level, so that the window in which spectroscopy is possible (which we refer to as the 'bias window') increases with increased magnitude of $V_{\mathrm{SD}}$. The current disappears again around zero bias, causing the tips of the diamonds to be shifted away from each other, because the drain (the hybrid NW) is gapped in this energy range. The measured current is positive when $V_{\mathrm{SD}}$ is below $0$~V, so in this configuration electrons flow into the device, and negative when $V_{\mathrm{SD}}$ is above $0$~V, corresponding to electrons flowing out of the device (or conversely holes flowing in). This demonstrates the action of the single QD levels as charge filters \cite{wang_nonlocal_2022}.
An additional feature is seen in Fig.~\ref{fig:datazerofield}(c) when compared to Fig.~\ref{fig:datazerofield}(b); this corresponds to the ABS resonance. The NW DOS information can be accessed more quantitatively by considering a $1$D line cut through a Coulomb diamond at fixed $V_\mathrm{SD}$, changing the energy of the QD by gating. The gate voltage scale is converted to energy using the gate lever arm, which is extracted from high-resolution differential conductance measurements of the Coulomb resonances by finding the slopes of both edges of the Coulomb diamonds, as shown in Fig.~\ref{fig:leverarm}. The slopes $m_1$ (red points) and $m_2$ (blue points) are combined to give the lever arm, $\alpha = 1/(m_1-m_2)$. For this set of resonances, $\alpha = 0.502$. 
The current measured is proportional to the DOS. Such $1$D measurements showing the NW DOS at zero field are shown in Figs.~\ref{fig:1dcuts0field}(a,~b) for the case of being on resonance and off resonance with the ABS in the gap, respectively, at the same $V_{\mathrm{P3}}$ values as  Fig.~\ref{fig:datazerofield}(b, c) as indicated by the color coding. The QD spectroscopy can be compared directly to QPC spectroscopy at corresponding $V_{\mathrm{P3}}$ values (Figs. \ref{fig:1dcuts0field}(c,d)). In the case with no ABS, both the QD and the QPC measurements show superconducting coherence peaks at $\sim \pm 290$ $\mu$eV. The QPC measurement further shows additional features at higher voltage bias, while the QD spectroscopy measurement is cut off by the value chosen for $V_{\mathrm{SD}}$ on the normal the lead ($0.4$ mV). When $V_{\mathrm{P3}}$ is adjusted so that an ABS comes down into the gap, the bound state energies can be read off using the QD spectrometer to be $\sim \pm 100$ $\mu$eV in agreement with the QPC measurement. 

\section{Spin resolved spectroscopy}\label{highfield}

The DOS of the NW can be measured using a single QD level at low field, but this does not provide any information that could not be obtained by tunneling spectroscopy with a QPC, which is readily accessible in these devices. The advantage of measuring through a QD level becomes apparent when one applies a finite magnetic field parallel to the NW, of a magnitude strong enough that the Zeeman splitting of the QD levels is greater than the desired energy window for the measurement. In this regime, the current that flows through a spin-polarized QD level is itself spin-polarized, so that measurements through levels of different spin polarization give spin-resolved DOS information. It is important to note that to interpret such measurements, one must keep in mind that there are multiple $g$-factors to be considered; that of the QD levels ($g_{\rm dot}$) and those of the system which is being probed via the QD level, in this case the hybrid NW system. 

\begin{figure*}[t]
\includegraphics[width=\textwidth]{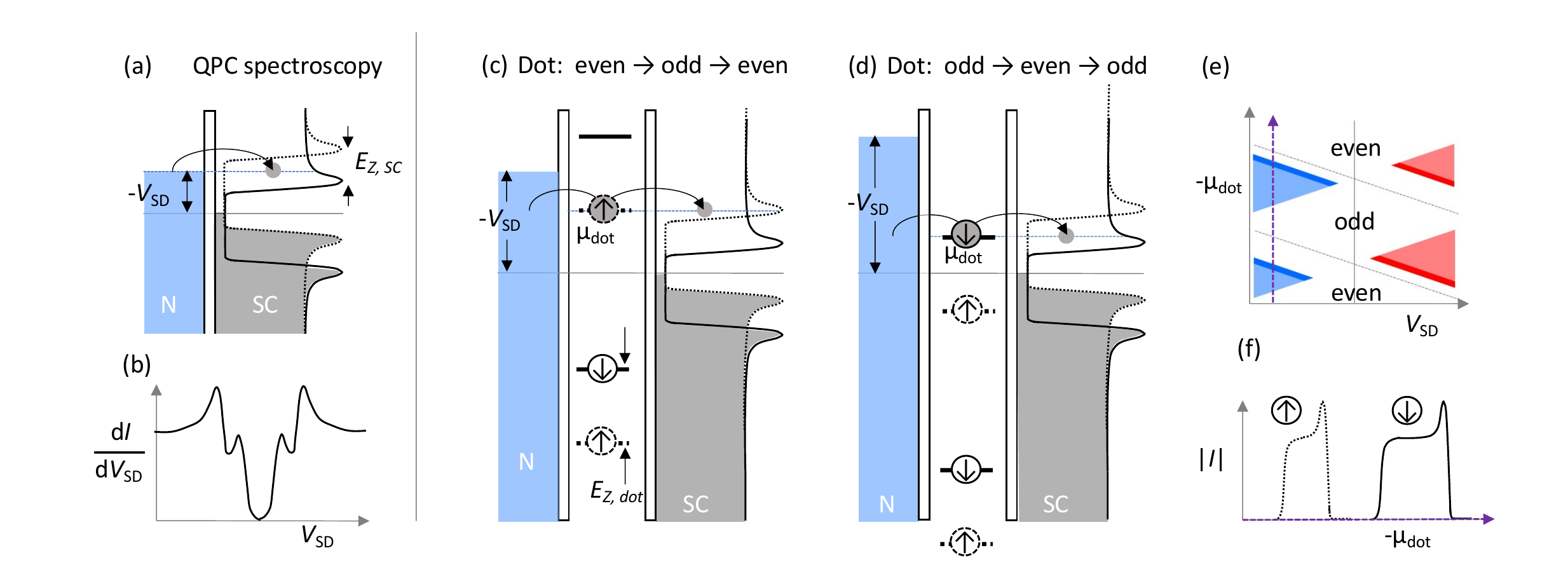}
\caption{\label{fig:finitetheory}Schematic illustration of the use of two consecutive QD levels for spin-resolved spectroscopy at finite magnetic field. (a) Sketch of the electrochemical potentials involved in conventional tunneling spectroscopy when the Al DOS is spin-split by $E_\mathrm{Z,SC}$. (b) The resulting differential conductance signal, sketched as a function of $V_\mathrm{SD}$. Spin-up and spin-down parts of the NW DOS are observed together. (c, d) Schematic of spin filtered tunneling current through a QD with levels which are split by the magnetic field, for even and odd filled ground states respectively. This results in a spin-down polarized current in the case of an even ground state, and a spin-up in the case of an odd ground state. (e) Sketch of the Coulomb diamond features for two consecutive levels expected as a function of $V_{\mathrm{SD}}$ for this configuration, with an asymmetry arising from the spin-splitting of the DOS and the spin polarized transport. (f) A slice through the two Coulomb diamonds at finite negative bias, showing separate spectroscopy of the spin-down and spin-up components of the NW DOS.}
\end{figure*} 

A schematic illustration of the use of spin-split QD levels as spin-selective spectrometers is given in Fig.~\ref{fig:finitetheory}. Here, only a superconducting gap is considered, without the added complication of any subgap features. When a field is applied, Cooper pairs keep their momentum pairing, but the opposite spin components of the pair have different energy \cite{meservey1994}. Since the $1e$ excited states remain separated in energy by $\Delta$ from the paired state, the coherence peaks appear at different energies for different spins, and the edges of the gap therefore split with some $g$-factor $g_{\mathrm{SC}}$. For bulk Al, $g =2$, but we can expect some modification of that in the hybrid system \cite{antipov_gfactor_2018}. The sketch of the tunneling process through a barrier is shown in Fig.~\ref{fig:finitetheory}(a), where one can see an electron tunneling from a normal lead at negative $V_\mathrm{SD}$ (not spin-polarized) into the NW DOS (spin-polarized). Since the normal lead and tunnel barrier are indifferent to spin, both spin-up and spin-down electrons can tunnel into the NW at a given $V_{\mathrm{SD}}$, and the resulting differential conductance signal is the total DOS (spin-up and spin-down components added together [Fig. \ref{fig:finitetheory}(b)]). The peaks which correspond to the spin-up and spin-down coherence peaks are still visible in the signal, but the two components are combined. To separate them, the tunneling current has to be spin filtered. This can be done by utilizing the QD as a spin selective barrier, tunneling through a single QD level which is spin-polarized, as in Figs.~\ref{fig:finitetheory}(c, d). 

As a magnetic field is applied, the levels of the QD will split with $g = g_{\rm dot}$, so the energy required to add a spin-up electron reduces in field, while the energy to add a spin-down electron to the same orbital increases. For the case of an even number of electrons $N$  on the QD, the most energetically favorable way to add another electron to the system is to load a spin-up electron into the next available level, so a spin-up current will predominantly flow. However, at low fields, the spin-down excited state is also accessible, so some transport through the first excited state will also be observed when both of the spin-split level components are within the voltage bias window. When the Zeeman splitting becomes larger than the selected bias window, the excited state is no longer available for transport, and the QD acts as a spin filter with no additional channels for the opposite spin. For the case shown in Fig. \ref{fig:finitetheory}(c), the excited state is already outside the bias window. For an odd number $N + 1$ electrons on the QD [(d)], the lower (spin-up) energy level is already filled by the $(N+1)^{th}$ electron, so the most energetically favorable transport option is to load a spin-down electron. The next excited state is much higher up in energy, so in the bias ranges which are used in this experiment the odd-even transition transport does not show any excited states. In this configuration, transport through two consecutive levels of the QD appears different at finite applied magnetic field; the even-odd transition filters spin-up electrons, but also shows transport through an excited state at lower fields, while the odd-even transition filters spin-down electrons. Figure \ref{fig:finitetheory}(e) shows a sketch of the current one can expect to measure through two such consecutive transitions, at a field where the Zeeman splitting of the QD is greater than the voltage bias range, $E_{Z, \mathrm{dot}} > eV_{\mathrm{SD}}$, so that no excited state transport is visible. A variation is expected in the size of the diamond tips, because the gap edge measured by one level (spin-down) is lower in energy than the edge measured by the other (spin-up). This can be visualized more directly by taking a 1D cut at finite bias through the two levels, as shown in (f). The leftmost edge of each resonance corresponds to the bias edge, where the QD level comes on resonance with the normal lead and sequential tunneling turns on. This feature is always at a fixed energy, as it is simply determined by the $V_\mathrm{SD}$ chosen for the measurement. This switching on is followed by a current that is proportional to the DOS; first a plateau above the gap energy, and then a peak in current, which corresponds to the coherence peak. Note that the distance between the bias edge and the coherence peak is different for the spin-up and spin-down resonances. This is because the spin-up and spin-down components of the DOS are resolved separately with the two different QD levels. The spin-up and spin-down components of the DOS are split, and the difference between the energy of their peaks is given by $E_\mathrm{Z,SC}/\alpha$.

\begin{figure}
\includegraphics[width=0.48\textwidth]{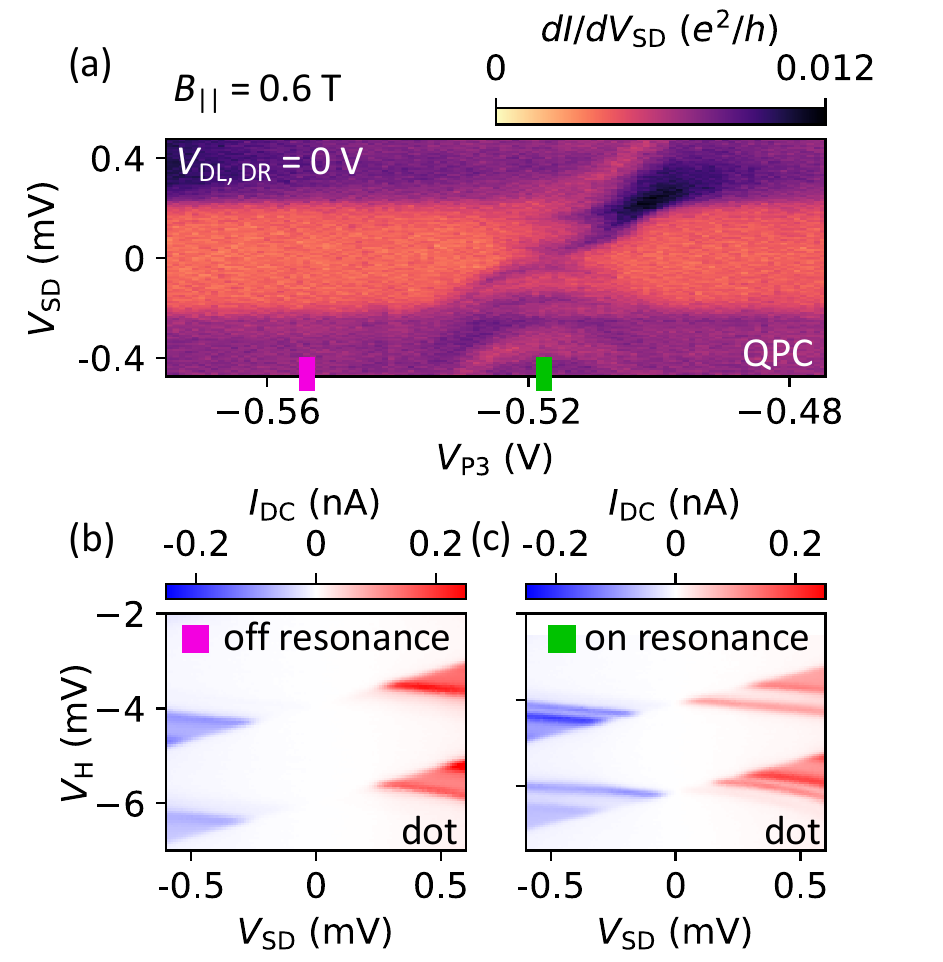}
\caption{\label{fig:datafinitefield}(a) QPC tunneling spectroscopy at $B_{||}=0.6$~T as a function of P3 gate voltage, $V_{\mathrm{P3}}$. The subgap state seen around $-0.52$~V is visibly spin split at this field value - four resonances are seen inside the gap instead of the two observed at low field. (b) Current measured with the QD formed, through two consecutive levels of the QD when $V_{\mathrm{P3}}$ is set far away from the subgap resonance (magenta). (c) Current through the same two levels of the QD with $V_{\mathrm{P3}}$ set to the location of the subgap state energy minimum (green).}
\end{figure}

We find that in device 1 the QD level even-odd pairs that are suitable for spectroscopy exhibit a $g$-factor of $\sim -8.5$, consistent with InAs confined in two dimensions \cite{smith_iii_g_1987}. This means that at an applied parallel field of $0.6$~T, the corresponding Zeeman splitting is $E_{Z, \mathrm{dot}}\sim 295$ $\mu$eV, so that features below that energy can already be spin-resolved by the QD spectrometer, but excited states of the QD still appear at higher biases. Measurements taken at this field value are shown in Fig.~\ref{fig:datafinitefield}. A tunnelling spectroscopy scan over NW potential (tuned by $V_{\mathrm{P3}}$) in Fig.~\ref{fig:datafinitefield}(a) shows a gap, reduced from the lower field value as expected, but with no visually resolvable splitting. The same subgap state seen in Fig.~\ref{fig:datazerofield} is also visible here, now split so that one component has moved towards zero energy while another has almost retreated into the continuum. Sequential tunneling current measured through the same two consecutive QD levels as in Fig.~\ref{fig:datazerofield} is shown in Figs.~\ref{fig:datafinitefield}(b, c) for $V_{\mathrm{P3}}$ values that bring the subgap state away from and onto resonance, respectively. An enhancement in current magnitude is seen in the bottom corner of the top left diamond tip in Fig. \ref{fig:datafinitefield}(b) (and equivalently in the top  corner of the bottom right diamond tip). These are signatures of the excited spin-down state, appearing at higher bias as expected. On resonance with the subgap state [Fig. \ref{fig:datafinitefield}(c)] the top right and bottom left diamond tips are notably larger than the other two. We interpret that this is due to their transport of spin-down electrons and corresponding spectroscopy of the spin-down part of the NW DOS, in this case measuring directly the part of the subgap state which moves towards zero energy. Additional resonances appear in the spin-up filtering diamond tips (top left and bottom right) due to spin-down transport via the spin-down excited state.

\begin{figure}
\includegraphics[width=0.48\textwidth]{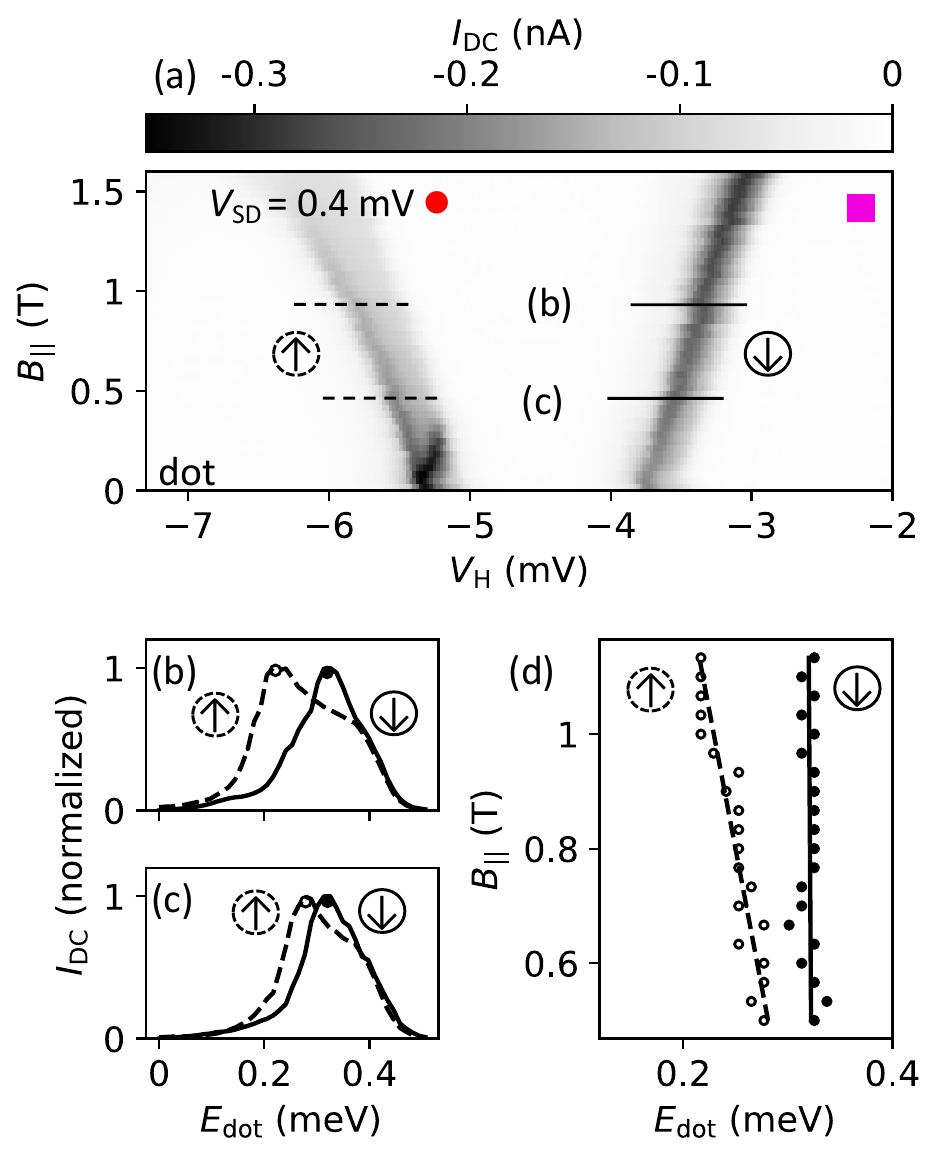}
\caption{\label{fig:gfactoral}Spin-resolved tunnelling current into NW with empty subgap. (a) Current through two consecutive levels of the weakly coupled QD at finite $V_{\mathrm{SD}}$ measured as a function of $B_{||}$. The levels move in opposite directions as the field is increased, indicating opposite spin. 1D slices through the right (solid line) and left (dashed line) resonances for 1 T (b) and 0.5~T (c). Horizontal axis converted from gate voltage to energy using lever arm. (d) Energy values of the coherence peaks of the spin-up and spin-down components of the DOS (from the left and right resonances) as a function of $B_{||}$ along with linear fits.}
\end{figure}

Measuring the dc current through the QD as a function of $V_\mathrm{SD}$ to see Coulomb diamonds is useful for confirming the behavior of excited states of the QD and extracting lever arms. However, to use a QD level as a tool for spin resolved spectroscopy and gain more insight to the DOS in the NW, it is sufficient to measure at a fixed $V_\mathrm{SD}$ on the normal lead. To examine directly the splitting of the superconducting gap in the absence of subgap states (off resonance) in field using the QD levels, $V_\mathrm{SD}$ is fixed at $+0.4$~mV, so that spectroscopic measurements can be taken via the two consecutive QD levels with a constant bias window. This can be thought of as taking a slice through Fig.~\ref{fig:datafinitefield}(b) at $V_{\mathrm{SD}}=+0.4$ mV, and then ramping the parallel magnetic field up from $0$ T. This measurement is shown in Fig.~\ref{fig:gfactoral}(a). Here, the result is a combined effect of two separate $g$-factors; the QD energy levels shifting in magnetic field, and the superconducting DOS evolving in field. The two consecutive QD states (transport through spin-up and spin-down ground states, respectively) move apart in field, and the previously discussed excited state can be seen as a high magnitude signal at low field in the left (spin-up) QD level. It splits rapidly in the opposite direction to the movement of the ground state. This motion of the QD levels does not provide any information about the NW DOS, and must be compensated for in the analysis. This compensation is possible because, as emphasized before, the bias edge of each resonance, where the current switches on because the QD level is on resonance with the normal lead, always corresponds to the same energy. So the position of the bias window in $V_{\mathrm{H}}$ space shifts in field, but the bias window remains the same. At each field value, the bias edge point is used as an anchor, so that when we convert the $V_{\mathrm{H}}$ axis from gate voltage to energy each one of the two resonances has a $0.4$ meV point, with respect to which a zero energy can be defined. 

Due to broadening of the QD levels, caused by finite temperature and finite coupling, the current does not go to zero instantaneously at the bias edge point and so the point itself is not perfectly defined. In this analysis, the placement of the bias edge for each resonance was determined by taking the $V_{\mathrm{H}}$ at which the current reached half of its peak value. This way, the method is standardized for each resonance, so relative energy values should be consistent with each other. The result of performing this analysis for each line of the measurement is that for each field value, one acquires two traces that are proportional to the DOS in the NW between $0$ and $0.4$ meV, one for the spin-up and one for the spin-down component of the NW DOS. This is shown in Figs.~\ref{fig:gfactoral}(b, c) for the field values of $1$ T and $0.5$ T respectively, with the normalized current traces for spin-up and spin-down plotted together in each case. At $0.5$~T there is a small splitting between the spin-up and spin-down coherence peaks, while at $1$~T the splitting is more significant. In Fig.~\ref{fig:gfactoral}(d) the peak energies extracted in a similar manner are plotted as a function of field, for values up to $1.1$~T (starting at $0.5$~T, so that the excited state is already outside the bias window). Linear fits to the splitting yield a the $g$-factor of $\sim 1.7$. This is slightly lower than $g_{Al} = 2$. This may be explained by hybridization at the interface between Al ($g_{\mathrm{Al}}=2$) and InAs ($g_{\mathrm{InAs}}=-15$) \cite{antipov_gfactor_2018}. Note that the spin-up peak splits more rapidly towards zero energy than spin-down splits away. This might be explained by the effect of spin orbit interaction \cite{meservey1994}. A similar analysis was performed for the negative bias side, where an electron current flows into the device instead of a hole current, showing a similar $g$-factor.

\begin{figure}
\includegraphics[width=0.48\textwidth]{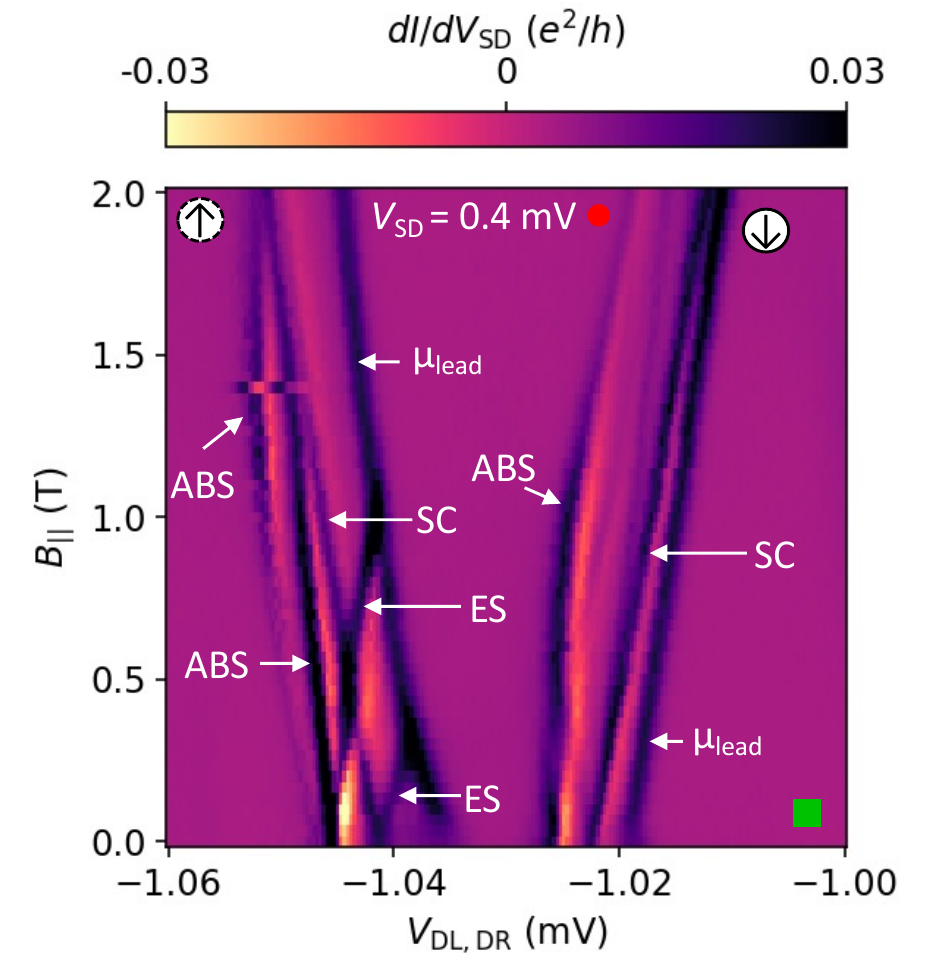}
\caption{\label{fig:dotfieldscan}Differential conductance through two consecutive levels of the weakly coupled QD at finite $V_{\mathrm{SD}}$ measured as a function of $B_{||}$, with $V_{\mathrm{P3}}$ set so that the subgap state in the NW is on resonance (indicated by the green square). Features corresponding to different spin components of the state (ABS) and the spin-split gap coherence peaks (SC) are indicated, as well as resonances arising from excited states of the QD itself (ES).}
\end{figure}

In the regime described above, one is able to separately measure the spin components of the spin-split superconducting gap in the NW, with no subgap features involved. If $V_{\mathrm{P3}}$ is adjusted, the spin-resolved field dependence of the previously shown subgap state can be investigated. Looking back at the tunneling spectroscopy measurement of this dependence in Fig.~\ref{fig:device}(d), it can be observed that the bound state splits in applied parallel magnetic field, with one component moving away from zero energy and merging into the continuum just above $1$~T, and the other moving towards zero energy before anticrossing at around $1$ T. By measuring via the two spin-selective QD levels, as with the gap splitting above, these different features of the state transport can be observed with spin resolution. For this purpose, it can be useful to record differential conductance as well as the dc current through the QD level. Although for sequential current measured through a weakly coupled QD level it is the dc current which is proportional to the DOS in the NW, a differential conductance measurement will (by definition) show a clear signal at points where the current undergoes a change as a function of $V_\mathrm{SD}$, so both peaks and regions of rapid change (such as the bias edge) are highlighted. Such a measurement is shown in Fig.~\ref{fig:dotfieldscan}, with $V_{\mathrm{P3}}$ set to the value at which the subgap state reaches a minimum at low field. Each bright resonance is labelled with the transport feature which it corresponds to, according to our interpretation. The rightmost resonances of both levels, labelled `$\mu_{\mathrm{lead}}$', correspond to the bias edge, where the QD level energy is on resonance with the normal lead, and the current switches on. The two resonances labelled `SC' correspond to the coherence peaks, for spin-up and spin-down respectively. Here, as before, it is important to make the distinction between the effect of the magnetic field on the NW DOS, which is being measured through the QD levels, and the movement of the levels themselves in field, which does not depend on the NW but purely on the QD. As in the analysis above, the splitting of the SC gap edges in field is not determined from the absolute movement of the `SC' resonances, but from their movement relative to their respective bias edge. The spin-up `SC' resonance moves away from the bias edge, towards zero energy, while the spin-down moves towards its respective bias edge, splitting away from zero energy. 

\begin{figure}
\includegraphics[width=0.48\textwidth]{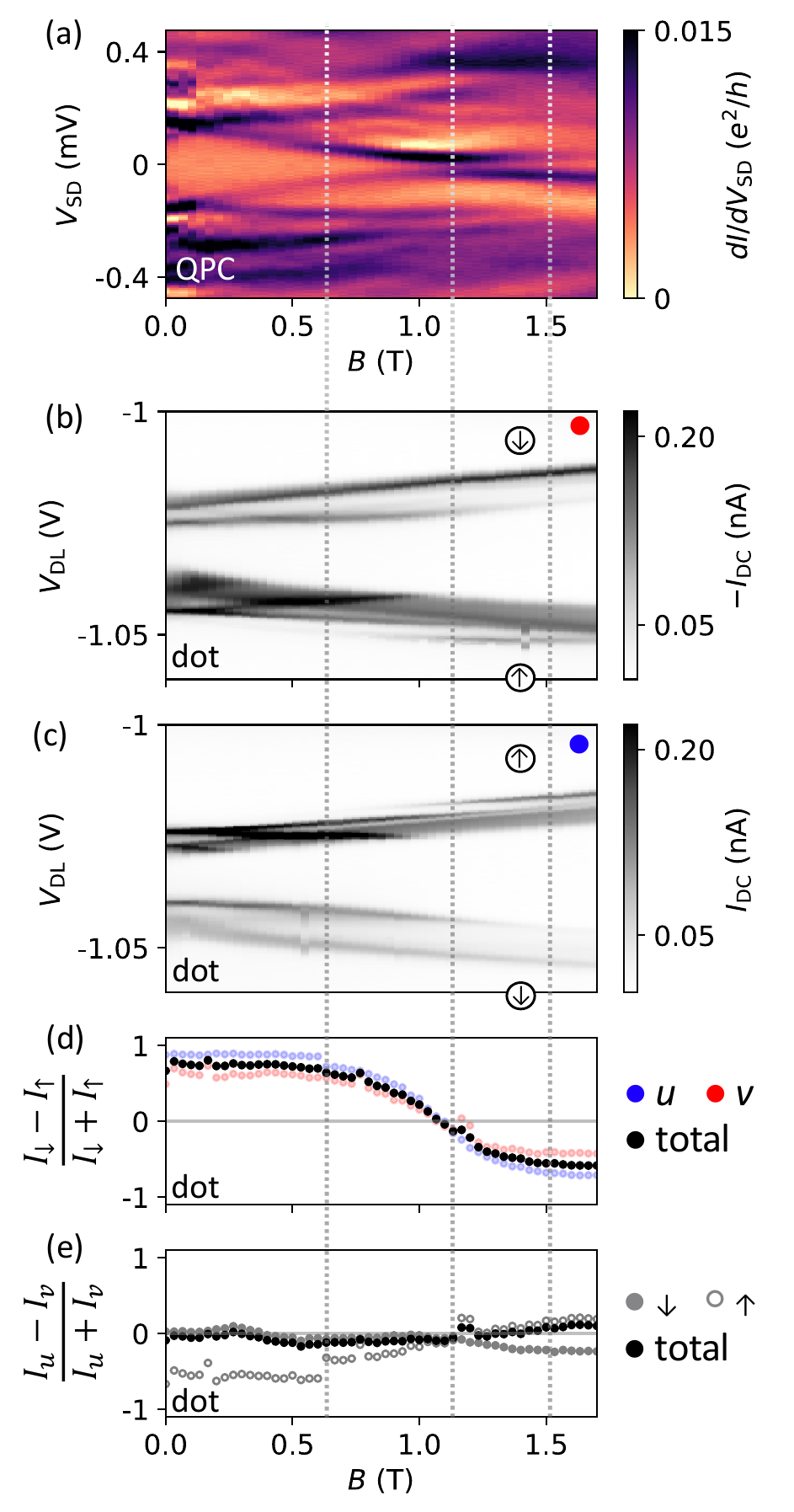}
\caption{\label{fig:spinresfield} Resolving the spin and charge character of a NW state while varying $B_{||}$. (a) Tunneling spectroscopy measurement of a localized NW state under gate P3 for increasing $B_{||}$. Gates forming the QD set to $0$~V. (b, c) Sequential tunnelling spectroscopy through consecutive levels of the QD, with $V_{\mathrm{SD}}$ fixed at $-0.4$~$\mathrm{m V}$ (b) and $+0.4$~$\mathrm{m V}$ (c). Current through the two QD levels is spin polarized, allowing spin-up and spin-down components of the NW DOS to be measured separately.  (d) Comparing the magnitudes of the tunneling current into the lowest energy state measured through the spin-up and spin-down resonances yields spin polarization, $S$, of the tunneling current, which crosses through zero around $1$ T, the field at which an anticrossing around zero bias is observed in the QPC measurement. (e) Comparing the magnitudes at positive and negative $V_{\mathrm{SD}}$ yields electron-hole polarization, 
which for this $V_{\mathrm{P3}}$ value stays close to zero throughout the measured field range.}
\end{figure}

Consider now the resonances caused by the NW state components labelled `ABS'. At lower field ($B_{||}<0.8$ T), a bright feature labelled `ABS' seen via the spin-down QD resonance moves away from the bias edge towards zero energy. This is in contrast to the spin-down `SC' component, which moves towards the bias edge. This observation suggests that the bound state and the gap edge have $g$-factors of opposite sign, a property which is not observable with standard tunneling spectroscopy. The `ABS' component, which splits in the opposite direction, away from zero energy, is observed via the spin-up resonance. Using a lever arm $\alpha= 0.043$, the two spin components of the ABS appear to split with a $g_{\mathrm{ABS}}\sim-2.25$ at low field. This is plausible, considering again that the $g$-factor of the hybrid system is renormalized by the hybridization between the Al and InAs, and that the effective $g$-factor for the bound state depends on the strength of this hybridization \cite{antipov_gfactor_2018}. Two resonances labelled `ES' are visible in transport via the spin-down excited state of the QD. As the field approaches $1$~T, the ABS feature which splits towards zero energy starts to fade in magnitude as measured via the spin-down level, and simultaneously appears via the spin-up resonance. After $1$~T it appears more brightly on the spin-up side than on the spin-down. This indicates a change in the ground state of the ABS component, as it goes from transporting primarily spin-down current to spin-up. Note that this change is not abrupt, it is a gradual transition.

\section{Spin and charge polarization of tunneling current}\label{spinres}

The transition between transport of spin-up to spin-down electrons via the bound state, which is resolved by measuring spin-up and spin-down current separately using the Zeeman split QD levels, can be quantified by defining a spin polarization of the transport through the state. Measuring at negative $V_\mathrm{SD}$ on the lead, the electron components of the transport current are accessed, so using two consecutive levels, one accesses separately the spin-up, electron component and the spin-down, electron component of the DOS. Similarly, by measuring at positive $V_\mathrm{SD}$, a hole current flows, and the spin-down, hole and spin-up, hole components are resolved. These four separate components are labelled explicitly in Figs.~\ref{fig:spinresfield}(b, c). We define a spin polarization $S$ by comparing the magnitude of the current into the state of interest as measured via the spin-up and spin-down components \cite{soulen_measuring_1998}. Separate polarization quantities can be extracted for the electron (u) and hole (v) measurements; 
\begin{equation}
    S_u = \frac{I_{\downarrow, u} - I_{\uparrow, u}}{I_{\downarrow, u} + I_{\uparrow, u}}
\end{equation}
\begin{equation}
    S_v = \frac{I_{\downarrow, v} - I_{\uparrow, v}}{I_{\downarrow, v} + I_{\uparrow, v}}.
\end{equation}
The electron and hole components can then be combined to define a total spin polarization
\begin{equation}
    S_{\rm{total}} = \frac{I_{\downarrow, u} - I_{\uparrow, u}+I_{\downarrow, v} - I_{\uparrow, v}}{I_{\downarrow, u} + I_{\uparrow, u}+I_{\downarrow, v} + I_{\uparrow, v}}.
\end{equation} 

\begin{figure}
\includegraphics[width=0.48\textwidth]{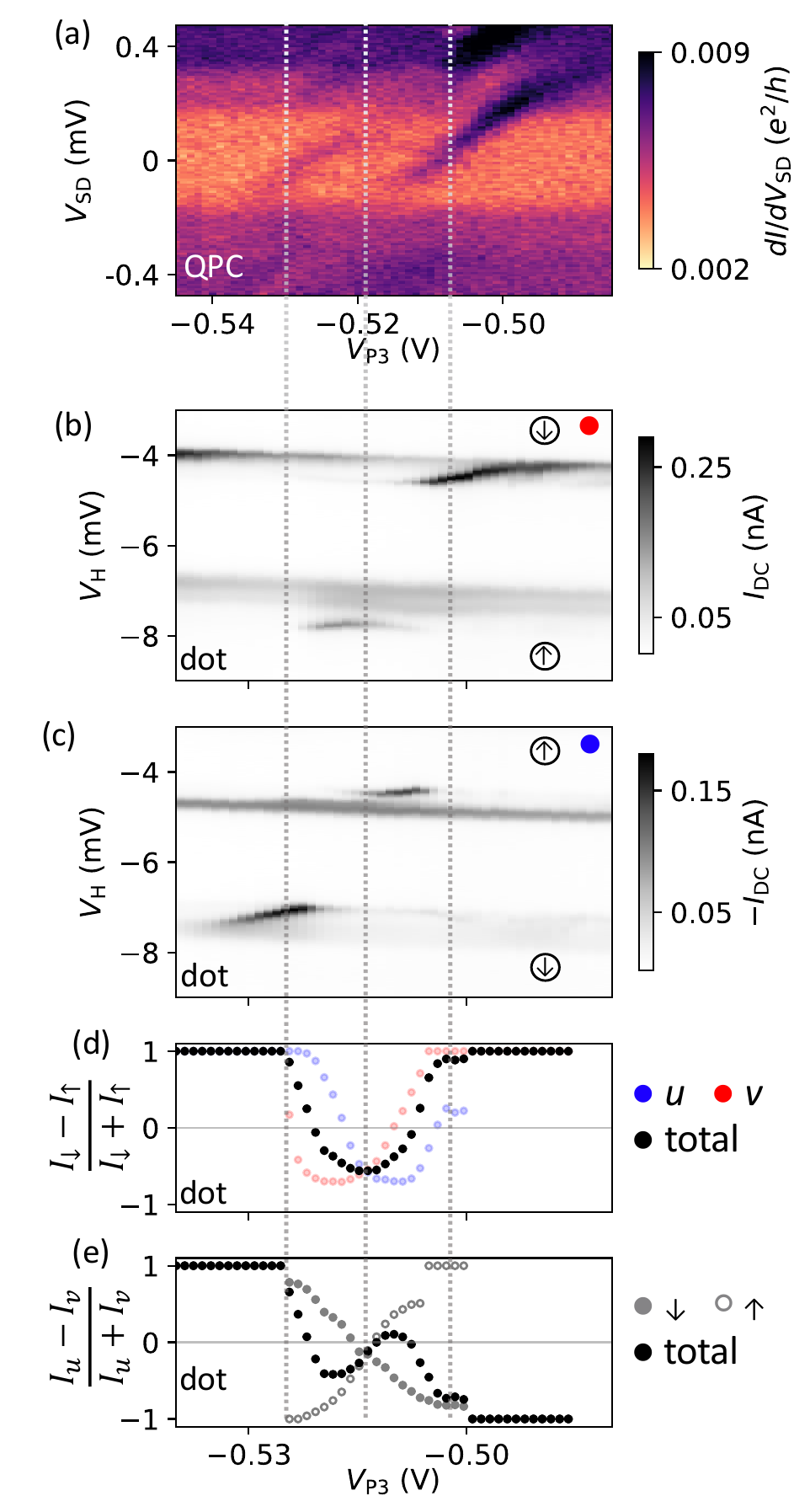}
\caption{\label{fig:spinrespotential}Resolving the spin and charge character of an ABS in the NW while varying gate voltage at a magnetic field of $1.4$ T. (a) Tunneling spectroscopy measurement of an ABS under the segment of NW tuned by gate P3 as a function of $V_{\mathrm{P}3}$. Gates forming the QD set to $0$~V. (b, c) The effect of the NW gate voltage measured using sequential tunnelling spectroscopy through consecutive levels of the QD, with $V_{\mathrm{SD}}$ at $-0.4$~$\mathrm{m V}$ (b) and $+0.4$~$\mathrm{m V}$ (c). The current through the two QD levels is spin polarized due to the Zeeman effect, so that the spin-up and spin-down components of the NW DOS are measured separately. (d) Spin polarisation, $S$, of the tunnelling current as a function of gate voltage. (e) Comparing the magnitudes at positive and negative $V_{\mathrm{SD}}$ leads to a measure of charge polarization $Q$ of the state.  }
\end{figure}

In a similar manner, a particle-hole polarization $Q$ can be defined by combining the relevant current magnitudes, yielding:
\begin{equation}
    Q_\downarrow = \frac{I_{\downarrow, u} - I_{\downarrow, v}}{I_{\downarrow, u} + I_{\downarrow, v}}.
\end{equation}
\begin{equation}
    Q_\uparrow = \frac{I_{\uparrow, u} - I_{\uparrow, v}}{I_{\uparrow, u} + I_{\uparrow, v}}.
\end{equation}
\begin{equation}
    Q_{total} = \frac{I_{\downarrow, u} + I_{\uparrow, u}-I_{\downarrow, v} - I_{\uparrow, v}}{I_{\downarrow, u} + I_{\uparrow, u}+I_{\downarrow, v} + I_{\uparrow, v}}.
\end{equation}

Using these definitions, the evolution of the spin and charge character of a bound state can be tracked with respect to a parameter like magnetic field or gate voltage. In Fig.~\ref{fig:spinresfield}, $V_{\mathrm{P3}}$ is set so that the previously investigated bound state is on resonance, and the four tunnelling current components ($I_{\downarrow, u}, I_{\uparrow, u}, I_{\downarrow, v}$ and $I_{\uparrow, v}$) are measured via two consecutive QD levels as before as $B_{||}$ is increased [Figs.~\ref{fig:spinresfield}(b, c)]. The spin and charge polarization are extracted from this data for the lowest energy ABS component, which is tracked using the \texttt{scipy} peak finder function. If a peak cannot be identified because it is below the noise level, the magnitude contribution is set to zero. The results of the extraction are shown in Fig.~\ref{fig:spinresfield}(d) for spin, including the separate electron and hole components and the total value, and in Fig.~\ref{fig:spinresfield}(e) for the total charge polarization. At low field, the current into the state is strongly spin-down polarized. In tunnelling spectroscopy (shown in Fig.~\ref{fig:spinresfield}(a) for comparison to the QD measurements) the state splits towards zero energy linearly, and in the QD measurements is observed almost exclusively through the spin-down filtering resonances. As the state approaches zero energy, an anticrossing is observed in tunnelling spectroscopy. This point, marked with the middle dashed line in Fig.~\ref{fig:spinresfield}, also marks the point where the currents measured through the spin-up and spin-down filtering QD levels and the lowest energy state are equal, leading to a net zero spin polarization.  

\begin{figure}
\includegraphics[width=0.48\textwidth]{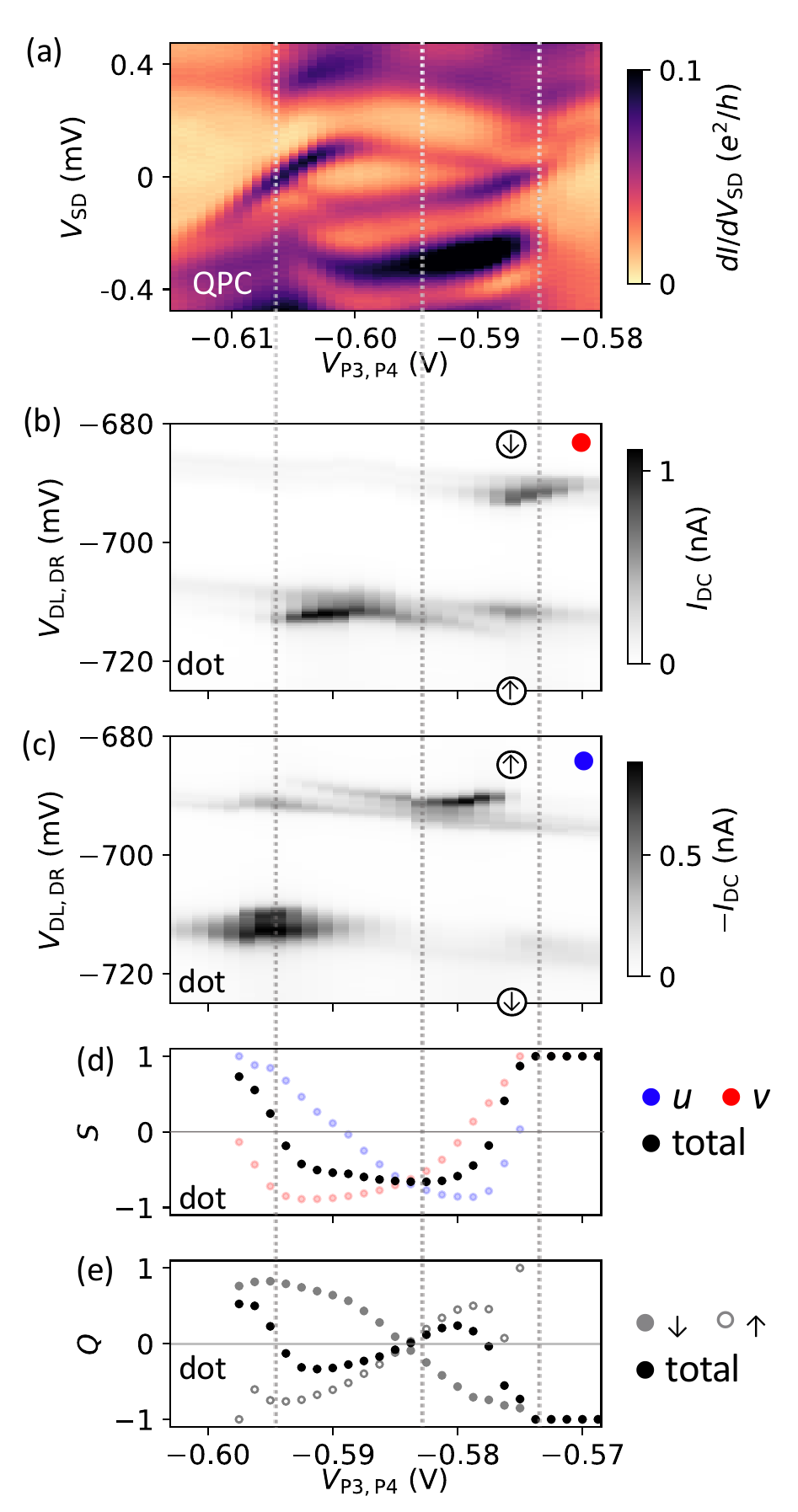}
\caption{\label{fig:spinrespotential2}Resolving the spin and charge character of an ABS in the NW in device 2 while varying the gate voltage at a magnetic field of 0.6 T. (a) Tunneling spectroscopy of an ABS under gate P3 for varying $V_{\mathrm{P}3}$. Gates that form the QD set to $0$~V. (b, c) the effect of the NW gate voltage variation measured instead using sequential tunnelling spectroscopy through consecutive levels of the QD, with $V_{\mathrm{SD}}$ fixed at $-400$~$\mathrm{\mu V}$ (b) and $+400$~$\mathrm{\mu V}$ (c). The current through the two QD levels is spin polarized, so that the spin-up and spin-down components of the NW DOS are measured separately. (d) Comparing the magnitudes of the current through the spin-up and spin-down resonances yields spin polarization, $S$, as a function of gate voltage. (e) Comparing the magnitudes at positive and negative $V_{\mathrm{SD}}$ yields charge polarization of the ABS. }
\end{figure}

Above the anticrossing,  the current measured through the spin-down filtering levels decays and the lowest energy state is mostly observed via the spin-up filtering level. Correspondingly, the spin polarization, having gone through zero at the point of anticrossing, switches to negative values (spin-up polarized). Our interpretation is that the spin polarization of the current reflects the spin polarization of the ABS. This crossing through zero is then consistent with a transition of the ABS in which the spin of the ground state switches in field \cite{lee_spin-resolved_2014,whiticar_parity_2021}. The charge polarization is also extracted (e); this appears to remain around zero for the entire field range, indicating that at the chosen gate voltage $V_{\mathrm{P3}}$ the DOS is equal parts electron and hole. 

This transition is further investigated by applying a fixed $B_{||}$ of $1.4$ T, 
above the field at which the anticrossing is observed, and changing the chemical potential by sweeping $V_{\mathrm{P3}}$. This measurement in shown for the same state as before in Fig.~\ref{fig:spinrespotential}, with a tunnelling spectroscopy measurement shown for comparison (a) and the four spin/charge components measured through the QD levels (b, c). For these data, the spin and charge polarization of the transport through the lowest-energy ABS are extracted in the same way as before, by peak-finding to track the energy of the state and taking the magnitude of the peak for each component $I_{\downarrow, u}, I_{\uparrow, u}, I_{\downarrow, v}$ and $I_{\uparrow, v}$.  In the tunnelling spectroscopy measurement, the state is observed to cross twice through zero energy, undergoing a characteristic singlet to doublet transition \cite{pillet_tunneling_2013, chang_tunneling_2013, whiticar_parity_2021, lee_spin-resolved_2014}.The switching of the ground state spin is directly seen from the spin polarization [Fig.~\ref{fig:spinrespotential}(d)]. The charge polarization dependence on the chemical potential is also nontrivial; this quantity crosses through zero three times, including once in the center of the doublet region. This is the chemical potential at which the magnetic field dependence was shown in Fig.~\ref{fig:spinresfield}. The charge polarization dependence is consistent with the Bardeen-Cooper-Schrieffer (BCS) charge quantity extracted from non-local conductance measurements of similar subgap states \cite{gramich_andreev_2017, schindele_nonlocal_2014, poschl_nonlocal_2022}. 

A similar data set, in which gate voltage is swept and the spin and charge polarization quantities of the transport through a local bound state are extracted from dc current measured via two consecutive QD levels, is shown in Fig.~\ref{fig:spinrespotential2}. This data is taken on a different device to the data shown in the rest of the paper, device 2, which is structurally similar to device 1. While the state under investigation looks quite different, exhibiting a much higher $g$-factor, the core features extracted from the data show a clear similarity to the observations from device 1. This is true for the general behavior in energy, as well as the behavior of the spin and electron-hole polarization. 

\section{Discussion}\label{discuss}
We have demonstrated the use of single QD levels to directly measure the DOS of a hybrid superconductor-semiconductor NW. For a QD in which the level spacing is larger than the superconducting gap $\Delta$ and the $g$-factor $|g_{\mathrm{dot}}| > |g_{\mathrm{SC}}|, |g_{\mathrm{ABS}}|$, Zeeman split QD levels of opposite spin character can be used to measure the density of states with spin and charge resolution. From these measurements, relative signs of $g$-factors are determined, and spin and charge polarizations extracted. 

Spin filtering using a laterally defined QD level tuned into an appropriate bias window has been suggested \cite{recher2000} and demonstrated \cite{hanson2003} before in the context of spin qubits, where the lifetime of an excited spin state was investigated. QD levels have also been used as spectrometers in the sense of reading out the numerical value of a superconducting gap \cite{junger_dot_2019}, and capacitive coupling considerations have been used to disentangle resonances in a Coulomb diamond caused by excited states of the QD itself from those which reflect the density of states in the leads \cite{thomas2021}. However, this work is the first to our knowledge to directly measure the evolution of the density of states of a hybrid system via a QD level, and to use Zeeman splitting of the levels to separately access spin-up and spin-down components of the density of states, and to resolve the relative sign of the $g$-factors of different spectroscopic features. Our spin polarization results are consistent with the physics of a singlet-to-doublet transition of an ABS \cite{lee_spin-resolved_2014}, as well as consistent with the BCS charge anticipated theoretically \cite{karsten_nl_spectroscopy} and extracted from non-local conductance measurements \cite{schindele_nonlocal_2014, denise_nl_gapclosing, poschl_nonlocal_2022}. The QDs used in this work are not few-electron QDs, as previously used for spin resolved tunnelling in 2DEG QDs \cite{hanson_bipolar_2004, hanson_spins_2007}. Instead, we use carefully selected levels of a many-electron QD which exhibit the desired filtering behavior in field, including splitting away from each other at low field, and the expected excited state behavior. This allows us to loosen the requirements on device design for future spin-filter QDs; it is not necessary to be able to deplete the QD fully to zero electrons, just to the point where there is a clear even-odd structure which can be associated with consecutive spin filling. 

Future work on the topic of subgap excitations in superconductor-semiconductor structures will benefit from this tool to separate the spin and charge components of the density of states, with the filtering properties coming as a very natural consequence of embedding a QD inside a tunnel probe. The deliberate definition of the QD in the design presented here has the added flexibility of allowing the QD to be turned off by setting all QD related gates to 0 V, so a direct comparison to standard tunneling spectroscopy is possible for any measurement. The gradual evolution seen in spin and charge polarization measurements hints at the strong spin orbit coupling present in the system \cite{szumniak_spin_2017}, and a combination of further experimental work with some theory could provide a new, direct method of extracting the spin-orbit coupling strength from spin and charge polarization quantities measured through a transition induced by field or chemical potential changes. Similar measurements could also be used to probe directly the inversion of the bulk bands at a phase transition point \cite{szumniak_spin_2017, chevallier_bulkband_2018}. In the current devices, we have so far only probed very local ABS features, which were accessible to only one probe at a time. However, similar structures have shown evidence of the presence of extended bound states \cite{PoschlPC2022}. Spin resolved measurements taken on both ends of a bound state simultaneously could provide even more information about the spin orbit coupling in these hybrid systems. 

\section{Acknowledgements}

We thank Serwan Asaad, Abhishek Banerjee, Asbjørn
Drachmann, David van Driel, Tom Dvir, William Lawrie, Magnus Ronne Lykkegard, Felix Passmann, Daniel
Sanchez, Saulius Vaitiek\.enas, and Frederik Knudsen Wolff
for input on experimental aspects.
We acknowledge support from the Danish National Research Foundation, Microsoft, and a research grant (Project 43951) from VILLUM FONDEN.

%\nocite{*}

\bibliography{references}% Produces the bibliography via BibTeX.

\begin{thebibliography}{46}%
\makeatletter
\providecommand \@ifxundefined [1]{%
 \@ifx{#1\undefined}
}%
\providecommand \@ifnum [1]{%
 \ifnum #1\expandafter \@firstoftwo
 \else \expandafter \@secondoftwo
 \fi
}%
\providecommand \@ifx [1]{%
 \ifx #1\expandafter \@firstoftwo
 \else \expandafter \@secondoftwo
 \fi
}%
\providecommand \natexlab [1]{#1}%
\providecommand \enquote  [1]{``#1''}%
\providecommand \bibnamefont  [1]{#1}%
\providecommand \bibfnamefont [1]{#1}%
\providecommand \citenamefont [1]{#1}%
\providecommand \href@noop [0]{\@secondoftwo}%
\providecommand \href [0]{\begingroup \@sanitize@url \@href}%
\providecommand \@href[1]{\@@startlink{#1}\@@href}%
\providecommand \@@href[1]{\endgroup#1\@@endlink}%
\providecommand \@sanitize@url [0]{\catcode `\\12\catcode `\$12\catcode
  `\&12\catcode `\#12\catcode `\^12\catcode `\_12\catcode `\%12\relax}%
\providecommand \@@startlink[1]{}%
\providecommand \@@endlink[0]{}%
\providecommand \url  [0]{\begingroup\@sanitize@url \@url }%
\providecommand \@url [1]{\endgroup\@href {#1}{\urlprefix }}%
\providecommand \urlprefix  [0]{URL }%
\providecommand \Eprint [0]{\href }%
\providecommand \doibase [0]{http://dx.doi.org/}%
\providecommand \selectlanguage [0]{\@gobble}%
\providecommand \bibinfo  [0]{\@secondoftwo}%
\providecommand \bibfield  [0]{\@secondoftwo}%
\providecommand \translation [1]{[#1]}%
\providecommand \BibitemOpen [0]{}%
\providecommand \bibitemStop [0]{}%
\providecommand \bibitemNoStop [0]{.\EOS\space}%
\providecommand \EOS [0]{\spacefactor3000\relax}%
\providecommand \BibitemShut  [1]{\csname bibitem#1\endcsname}%
\let\auto@bib@innerbib\@empty
%</preamble>
\bibitem [{\citenamefont {Kroger}\ \emph {et~al.}(1989)\citenamefont {Kroger},
  \citenamefont {Hilbert}, \citenamefont {Gibson}, \citenamefont {Ghoshal},\
  and\ \citenamefont {Smith}}]{kroger_superconductor-semiconductor_1989}%
  \BibitemOpen
  \bibfield  {author} {\bibinfo {author} {\bibfnamefont {H.}~\bibnamefont
  {Kroger}}, \bibinfo {author} {\bibfnamefont {C.}~\bibnamefont {Hilbert}},
  \bibinfo {author} {\bibfnamefont {D.}~\bibnamefont {Gibson}}, \bibinfo
  {author} {\bibfnamefont {U.}~\bibnamefont {Ghoshal}}, \ and\ \bibinfo
  {author} {\bibfnamefont {L.}~\bibnamefont {Smith}},\ }\href {\doibase
  10.1109/5.34127} {\bibfield  {journal} {\bibinfo  {journal} {Proceedings of
  the IEEE}\ }\textbf {\bibinfo {volume} {77}},\ \bibinfo {pages} {1287}
  (\bibinfo {year} {1989})}\BibitemShut {NoStop}%
\bibitem [{\citenamefont {Krogstrup}\ \emph {et~al.}(2015)\citenamefont
  {Krogstrup}, \citenamefont {Ziino}, \citenamefont {Chang}, \citenamefont
  {Albrecht}, \citenamefont {Madsen}, \citenamefont {Johnson}, \citenamefont
  {Nygård}, \citenamefont {Marcus},\ and\ \citenamefont
  {Jespersen}}]{krogstrup_epitaxy_2015}%
  \BibitemOpen
  \bibfield  {author} {\bibinfo {author} {\bibfnamefont {P.}~\bibnamefont
  {Krogstrup}}, \bibinfo {author} {\bibfnamefont {N.~L.~B.}\ \bibnamefont
  {Ziino}}, \bibinfo {author} {\bibfnamefont {W.}~\bibnamefont {Chang}},
  \bibinfo {author} {\bibfnamefont {S.~M.}\ \bibnamefont {Albrecht}}, \bibinfo
  {author} {\bibfnamefont {M.~H.}\ \bibnamefont {Madsen}}, \bibinfo {author}
  {\bibfnamefont {E.}~\bibnamefont {Johnson}}, \bibinfo {author} {\bibfnamefont
  {J.}~\bibnamefont {Nygård}}, \bibinfo {author} {\bibfnamefont {C.~M.}\
  \bibnamefont {Marcus}}, \ and\ \bibinfo {author} {\bibfnamefont {T.~S.}\
  \bibnamefont {Jespersen}},\ }\href {\doibase 10.1038/nmat4176} {\bibfield
  {journal} {\bibinfo  {journal} {Nature Materials}\ }\textbf {\bibinfo
  {volume} {14}},\ \bibinfo {pages} {400} (\bibinfo {year} {2015})}\BibitemShut
  {NoStop}%
\bibitem [{\citenamefont {Shabani}\ \emph {et~al.}(2016)\citenamefont
  {Shabani}, \citenamefont {Kjaergaard}, \citenamefont {Suominen},
  \citenamefont {Kim}, \citenamefont {Nichele}, \citenamefont {Pakrouski},
  \citenamefont {Stankevic}, \citenamefont {Lutchyn}, \citenamefont
  {Krogstrup}, \citenamefont {Feidenhans'l}, \citenamefont {Kraemer},
  \citenamefont {Nayak}, \citenamefont {Troyer}, \citenamefont {Marcus},\ and\
  \citenamefont {Palmstrøm}}]{shabani_two-dimensional_2016}%
  \BibitemOpen
  \bibfield  {author} {\bibinfo {author} {\bibfnamefont {J.}~\bibnamefont
  {Shabani}}, \bibinfo {author} {\bibfnamefont {M.}~\bibnamefont {Kjaergaard}},
  \bibinfo {author} {\bibfnamefont {H.~J.}\ \bibnamefont {Suominen}}, \bibinfo
  {author} {\bibfnamefont {Y.}~\bibnamefont {Kim}}, \bibinfo {author}
  {\bibfnamefont {F.}~\bibnamefont {Nichele}}, \bibinfo {author} {\bibfnamefont
  {K.}~\bibnamefont {Pakrouski}}, \bibinfo {author} {\bibfnamefont
  {T.}~\bibnamefont {Stankevic}}, \bibinfo {author} {\bibfnamefont {R.~M.}\
  \bibnamefont {Lutchyn}}, \bibinfo {author} {\bibfnamefont {P.}~\bibnamefont
  {Krogstrup}}, \bibinfo {author} {\bibfnamefont {R.}~\bibnamefont
  {Feidenhans'l}}, \bibinfo {author} {\bibfnamefont {S.}~\bibnamefont
  {Kraemer}}, \bibinfo {author} {\bibfnamefont {C.}~\bibnamefont {Nayak}},
  \bibinfo {author} {\bibfnamefont {M.}~\bibnamefont {Troyer}}, \bibinfo
  {author} {\bibfnamefont {C.~M.}\ \bibnamefont {Marcus}}, \ and\ \bibinfo
  {author} {\bibfnamefont {C.~J.}\ \bibnamefont {Palmstrøm}},\ }\href
  {\doibase 10.1103/PhysRevB.93.155402} {\bibfield  {journal} {\bibinfo
  {journal} {Phys. Rev. B}\ }\textbf {\bibinfo {volume} {93}},\ \bibinfo
  {pages} {155402} (\bibinfo {year} {2016})}\BibitemShut {NoStop}%
\bibitem [{\citenamefont {Takayanagi}\ and\ \citenamefont
  {Kawakami}(1985)}]{takayanagi_superconducting_1985}%
  \BibitemOpen
  \bibfield  {author} {\bibinfo {author} {\bibfnamefont {H.}~\bibnamefont
  {Takayanagi}}\ and\ \bibinfo {author} {\bibfnamefont {T.}~\bibnamefont
  {Kawakami}},\ }\href {\doibase 10.1103/PhysRevLett.54.2449} {\bibfield
  {journal} {\bibinfo  {journal} {Physical Review Letters}\ }\textbf {\bibinfo
  {volume} {54}},\ \bibinfo {pages} {2449} (\bibinfo {year}
  {1985})}\BibitemShut {NoStop}%
\bibitem [{\citenamefont {Beenakker}(1992)}]{beenakker_quantum_1992}%
  \BibitemOpen
  \bibfield  {author} {\bibinfo {author} {\bibfnamefont {C.~W.~J.}\
  \bibnamefont {Beenakker}},\ }\href {\doibase 10.1103/PhysRevB.46.12841}
  {\bibfield  {journal} {\bibinfo  {journal} {Physical Review B}\ }\textbf
  {\bibinfo {volume} {46}},\ \bibinfo {pages} {12841} (\bibinfo {year}
  {1992})}\BibitemShut {NoStop}%
\bibitem [{\citenamefont {Schapers}\ and\ \citenamefont
  {Schäpers}(2001)}]{schapers_proximity_2001}%
  \BibitemOpen
  \bibfield  {author} {\bibinfo {author} {\bibfnamefont {T.}~\bibnamefont
  {Schapers}}\ and\ \bibinfo {author} {\bibfnamefont {T.}~\bibnamefont
  {Schäpers}},\ }\href@noop {} {\emph {\bibinfo {title}
  {Superconductor/{Semiconductor} {Junctions}}}}\ (\bibinfo  {publisher}
  {Springer Science \& Business Media},\ \bibinfo {year} {2001})\BibitemShut
  {NoStop}%
\bibitem [{\citenamefont {Chrestin}\ \emph {et~al.}(1997)\citenamefont
  {Chrestin}, \citenamefont {Matsuyama},\ and\ \citenamefont
  {Merkt}}]{chrestin_soc_1997}%
  \BibitemOpen
  \bibfield  {author} {\bibinfo {author} {\bibfnamefont {A.}~\bibnamefont
  {Chrestin}}, \bibinfo {author} {\bibfnamefont {T.}~\bibnamefont {Matsuyama}},
  \ and\ \bibinfo {author} {\bibfnamefont {U.}~\bibnamefont {Merkt}},\ }\href
  {\doibase 10.1103/PhysRevB.55.8457} {\bibfield  {journal} {\bibinfo
  {journal} {Physical Review B}\ }\textbf {\bibinfo {volume} {55}},\ \bibinfo
  {pages} {8457} (\bibinfo {year} {1997})}\BibitemShut {NoStop}%
\bibitem [{\citenamefont {Fasth}\ \emph {et~al.}(2007)\citenamefont {Fasth},
  \citenamefont {Fuhrer}, \citenamefont {Samuelson}, \citenamefont {Golovach},\
  and\ \citenamefont {Loss}}]{fasth_direct_2007}%
  \BibitemOpen
  \bibfield  {author} {\bibinfo {author} {\bibfnamefont {C.}~\bibnamefont
  {Fasth}}, \bibinfo {author} {\bibfnamefont {A.}~\bibnamefont {Fuhrer}},
  \bibinfo {author} {\bibfnamefont {L.}~\bibnamefont {Samuelson}}, \bibinfo
  {author} {\bibfnamefont {V.~N.}\ \bibnamefont {Golovach}}, \ and\ \bibinfo
  {author} {\bibfnamefont {D.}~\bibnamefont {Loss}},\ }\href {\doibase
  10.1103/PhysRevLett.98.266801} {\bibfield  {journal} {\bibinfo  {journal}
  {Physical Review Letters}\ }\textbf {\bibinfo {volume} {98}},\ \bibinfo
  {pages} {266801} (\bibinfo {year} {2007})}\BibitemShut {NoStop}%
\bibitem [{\citenamefont {O’Connell~Yuan}\ \emph {et~al.}(2021)\citenamefont
  {O’Connell~Yuan}, \citenamefont {Wickramasinghe}, \citenamefont
  {Strickland}, \citenamefont {Dartiailh}, \citenamefont {Sardashti},
  \citenamefont {Hatefipour},\ and\ \citenamefont
  {Shabani}}]{oconnell_yuan_epitaxial_2021}%
  \BibitemOpen
  \bibfield  {author} {\bibinfo {author} {\bibfnamefont {J.}~\bibnamefont
  {O’Connell~Yuan}}, \bibinfo {author} {\bibfnamefont {K.~S.}\ \bibnamefont
  {Wickramasinghe}}, \bibinfo {author} {\bibfnamefont {W.~M.}\ \bibnamefont
  {Strickland}}, \bibinfo {author} {\bibfnamefont {M.~C.}\ \bibnamefont
  {Dartiailh}}, \bibinfo {author} {\bibfnamefont {K.}~\bibnamefont
  {Sardashti}}, \bibinfo {author} {\bibfnamefont {M.}~\bibnamefont
  {Hatefipour}}, \ and\ \bibinfo {author} {\bibfnamefont {J.}~\bibnamefont
  {Shabani}},\ }\href {\doibase 10.1116/6.0000918} {\bibfield  {journal}
  {\bibinfo  {journal} {Journal of Vacuum Science \& Technology A}\ }\textbf
  {\bibinfo {volume} {39}},\ \bibinfo {pages} {033407} (\bibinfo {year}
  {2021})}\BibitemShut {NoStop}%
\bibitem [{\citenamefont {Kjaergaard}\ \emph {et~al.}(2017)\citenamefont
  {Kjaergaard}, \citenamefont {Suominen}, \citenamefont {Nowak}, \citenamefont
  {Akhmerov}, \citenamefont {Shabani}, \citenamefont {Palmstrøm},
  \citenamefont {Nichele},\ and\ \citenamefont
  {Marcus}}]{kjaergaard_transparent_2017}%
  \BibitemOpen
  \bibfield  {author} {\bibinfo {author} {\bibfnamefont {M.}~\bibnamefont
  {Kjaergaard}}, \bibinfo {author} {\bibfnamefont {H.}~\bibnamefont
  {Suominen}}, \bibinfo {author} {\bibfnamefont {M.}~\bibnamefont {Nowak}},
  \bibinfo {author} {\bibfnamefont {A.}~\bibnamefont {Akhmerov}}, \bibinfo
  {author} {\bibfnamefont {J.}~\bibnamefont {Shabani}}, \bibinfo {author}
  {\bibfnamefont {C.}~\bibnamefont {Palmstrøm}}, \bibinfo {author}
  {\bibfnamefont {F.}~\bibnamefont {Nichele}}, \ and\ \bibinfo {author}
  {\bibfnamefont {C.}~\bibnamefont {Marcus}},\ }\href {\doibase
  10.1103/PhysRevApplied.7.034029} {\bibfield  {journal} {\bibinfo  {journal}
  {Phys. Rev. Appl.}\ }\textbf {\bibinfo {volume} {7}},\ \bibinfo {pages}
  {034029} (\bibinfo {year} {2017})}\BibitemShut {NoStop}%
\bibitem [{\citenamefont {Balatsky}\ \emph {et~al.}(2006)\citenamefont
  {Balatsky}, \citenamefont {Vekhter},\ and\ \citenamefont
  {Zhu}}]{balatsky_impurity-induced_2006}%
  \BibitemOpen
  \bibfield  {author} {\bibinfo {author} {\bibfnamefont {A.~V.}\ \bibnamefont
  {Balatsky}}, \bibinfo {author} {\bibfnamefont {I.}~\bibnamefont {Vekhter}}, \
  and\ \bibinfo {author} {\bibfnamefont {J.-X.}\ \bibnamefont {Zhu}},\ }\href
  {\doibase 10.1103/RevModPhys.78.373} {\bibfield  {journal} {\bibinfo
  {journal} {Reviews of Modern Physics}\ }\textbf {\bibinfo {volume} {78}},\
  \bibinfo {pages} {373} (\bibinfo {year} {2006})}\BibitemShut {NoStop}%
\bibitem [{\citenamefont {Whiticar}\ \emph {et~al.}(2021)\citenamefont
  {Whiticar}, \citenamefont {Fornieri}, \citenamefont {Banerjee}, \citenamefont
  {Drachmann}, \citenamefont {Gronin}, \citenamefont {Gardner}, \citenamefont
  {Lindemann}, \citenamefont {Manfra},\ and\ \citenamefont
  {Marcus}}]{whiticar_parity_2021}%
  \BibitemOpen
  \bibfield  {author} {\bibinfo {author} {\bibfnamefont {A.~M.}\ \bibnamefont
  {Whiticar}}, \bibinfo {author} {\bibfnamefont {A.}~\bibnamefont {Fornieri}},
  \bibinfo {author} {\bibfnamefont {A.}~\bibnamefont {Banerjee}}, \bibinfo
  {author} {\bibfnamefont {A.~C.~C.}\ \bibnamefont {Drachmann}}, \bibinfo
  {author} {\bibfnamefont {S.}~\bibnamefont {Gronin}}, \bibinfo {author}
  {\bibfnamefont {G.~C.}\ \bibnamefont {Gardner}}, \bibinfo {author}
  {\bibfnamefont {T.}~\bibnamefont {Lindemann}}, \bibinfo {author}
  {\bibfnamefont {M.~J.}\ \bibnamefont {Manfra}}, \ and\ \bibinfo {author}
  {\bibfnamefont {C.~M.}\ \bibnamefont {Marcus}},\ }\href {\doibase
  10.1103/PhysRevB.103.245308} {\bibfield  {journal} {\bibinfo  {journal}
  {Phys. Rev. B}\ }\textbf {\bibinfo {volume} {103}},\ \bibinfo {pages}
  {245308} (\bibinfo {year} {2021})}\BibitemShut {NoStop}%
\bibitem [{\citenamefont {Suominen}\ \emph {et~al.}(2017)\citenamefont
  {Suominen}, \citenamefont {Kjaergaard}, \citenamefont {Hamilton},
  \citenamefont {Shabani}, \citenamefont {Palmstrøm}, \citenamefont {Marcus},\
  and\ \citenamefont {Nichele}}]{henri_lead}%
  \BibitemOpen
  \bibfield  {author} {\bibinfo {author} {\bibfnamefont {H.}~\bibnamefont
  {Suominen}}, \bibinfo {author} {\bibfnamefont {M.}~\bibnamefont
  {Kjaergaard}}, \bibinfo {author} {\bibfnamefont {A.}~\bibnamefont
  {Hamilton}}, \bibinfo {author} {\bibfnamefont {J.}~\bibnamefont {Shabani}},
  \bibinfo {author} {\bibfnamefont {C.}~\bibnamefont {Palmstrøm}}, \bibinfo
  {author} {\bibfnamefont {C.}~\bibnamefont {Marcus}}, \ and\ \bibinfo {author}
  {\bibfnamefont {F.}~\bibnamefont {Nichele}},\ }\href {\doibase
  10.1103/PhysRevLett.119.176805} {\bibfield  {journal} {\bibinfo  {journal}
  {Phys. Rev. Lett.}\ }\textbf {\bibinfo {volume} {119}},\ \bibinfo {pages}
  {176805} (\bibinfo {year} {2017})}\BibitemShut {NoStop}%
\bibitem [{\citenamefont {Whiticar}\ \emph {et~al.}(2020)\citenamefont
  {Whiticar}, \citenamefont {Fornieri}, \citenamefont {O’Farrell},
  \citenamefont {Drachmann}, \citenamefont {Wang}, \citenamefont {Thomas},
  \citenamefont {Gronin}, \citenamefont {Kallaher}, \citenamefont {Gardner},
  \citenamefont {Manfra}, \citenamefont {Marcus},\ and\ \citenamefont
  {Nichele}}]{whiticar_coherent_2020}%
  \BibitemOpen
  \bibfield  {author} {\bibinfo {author} {\bibfnamefont {A.~M.}\ \bibnamefont
  {Whiticar}}, \bibinfo {author} {\bibfnamefont {A.}~\bibnamefont {Fornieri}},
  \bibinfo {author} {\bibfnamefont {E.~C.~T.}\ \bibnamefont {O’Farrell}},
  \bibinfo {author} {\bibfnamefont {A.~C.~C.}\ \bibnamefont {Drachmann}},
  \bibinfo {author} {\bibfnamefont {T.}~\bibnamefont {Wang}}, \bibinfo {author}
  {\bibfnamefont {C.}~\bibnamefont {Thomas}}, \bibinfo {author} {\bibfnamefont
  {S.}~\bibnamefont {Gronin}}, \bibinfo {author} {\bibfnamefont
  {R.}~\bibnamefont {Kallaher}}, \bibinfo {author} {\bibfnamefont {G.~C.}\
  \bibnamefont {Gardner}}, \bibinfo {author} {\bibfnamefont {M.~J.}\
  \bibnamefont {Manfra}}, \bibinfo {author} {\bibfnamefont {C.~M.}\
  \bibnamefont {Marcus}}, \ and\ \bibinfo {author} {\bibfnamefont
  {F.}~\bibnamefont {Nichele}},\ }\href {\doibase 10.1038/s41467-020-16988-x}
  {\bibfield  {journal} {\bibinfo  {journal} {Nat.~Comm.}\ }\textbf {\bibinfo
  {volume} {11}},\ \bibinfo {pages} {3212} (\bibinfo {year}
  {2020})}\BibitemShut {NoStop}%
\bibitem [{\citenamefont {Nichele}\ \emph {et~al.}(2017)\citenamefont
  {Nichele}, \citenamefont {Drachmann}, \citenamefont {Whiticar}, \citenamefont
  {O’Farrell}, \citenamefont {Suominen}, \citenamefont {Fornieri},
  \citenamefont {Wang}, \citenamefont {Gardner}, \citenamefont {Thomas},
  \citenamefont {Hatke}, \citenamefont {Krogstrup}, \citenamefont {Manfra},
  \citenamefont {Flensberg},\ and\ \citenamefont
  {Marcus}}]{nichele_scaling_2017}%
  \BibitemOpen
  \bibfield  {author} {\bibinfo {author} {\bibfnamefont {F.}~\bibnamefont
  {Nichele}}, \bibinfo {author} {\bibfnamefont {A.~C.}\ \bibnamefont
  {Drachmann}}, \bibinfo {author} {\bibfnamefont {A.~M.}\ \bibnamefont
  {Whiticar}}, \bibinfo {author} {\bibfnamefont {E.~C.}\ \bibnamefont
  {O’Farrell}}, \bibinfo {author} {\bibfnamefont {H.~J.}\ \bibnamefont
  {Suominen}}, \bibinfo {author} {\bibfnamefont {A.}~\bibnamefont {Fornieri}},
  \bibinfo {author} {\bibfnamefont {T.}~\bibnamefont {Wang}}, \bibinfo {author}
  {\bibfnamefont {G.~C.}\ \bibnamefont {Gardner}}, \bibinfo {author}
  {\bibfnamefont {C.}~\bibnamefont {Thomas}}, \bibinfo {author} {\bibfnamefont
  {A.~T.}\ \bibnamefont {Hatke}}, \bibinfo {author} {\bibfnamefont
  {P.}~\bibnamefont {Krogstrup}}, \bibinfo {author} {\bibfnamefont {M.~J.}\
  \bibnamefont {Manfra}}, \bibinfo {author} {\bibfnamefont {K.}~\bibnamefont
  {Flensberg}}, \ and\ \bibinfo {author} {\bibfnamefont {C.~M.}\ \bibnamefont
  {Marcus}},\ }\href {\doibase 10.1103/PhysRevLett.119.136803} {\bibfield
  {journal} {\bibinfo  {journal} {Phys. Rev. Lett.}\ }\textbf {\bibinfo
  {volume} {119}},\ \bibinfo {pages} {136803} (\bibinfo {year}
  {2017})}\BibitemShut {NoStop}%
\bibitem [{\citenamefont {O’Farrell}\ \emph {et~al.}(2018)\citenamefont
  {O’Farrell}, \citenamefont {Drachmann}, \citenamefont {Hell}, \citenamefont
  {Fornieri}, \citenamefont {Whiticar}, \citenamefont {Hansen}, \citenamefont
  {Gronin}, \citenamefont {Gardner}, \citenamefont {Thomas}, \citenamefont
  {Manfra}, \citenamefont {Flensberg}, \citenamefont {Marcus},\ and\
  \citenamefont {Nichele}}]{ofarrell_hybridization_2018}%
  \BibitemOpen
  \bibfield  {author} {\bibinfo {author} {\bibfnamefont {E.}~\bibnamefont
  {O’Farrell}}, \bibinfo {author} {\bibfnamefont {A.}~\bibnamefont
  {Drachmann}}, \bibinfo {author} {\bibfnamefont {M.}~\bibnamefont {Hell}},
  \bibinfo {author} {\bibfnamefont {A.}~\bibnamefont {Fornieri}}, \bibinfo
  {author} {\bibfnamefont {A.}~\bibnamefont {Whiticar}}, \bibinfo {author}
  {\bibfnamefont {E.}~\bibnamefont {Hansen}}, \bibinfo {author} {\bibfnamefont
  {S.}~\bibnamefont {Gronin}}, \bibinfo {author} {\bibfnamefont
  {G.}~\bibnamefont {Gardner}}, \bibinfo {author} {\bibfnamefont
  {C.}~\bibnamefont {Thomas}}, \bibinfo {author} {\bibfnamefont
  {M.}~\bibnamefont {Manfra}}, \bibinfo {author} {\bibfnamefont
  {K.}~\bibnamefont {Flensberg}}, \bibinfo {author} {\bibfnamefont
  {C.}~\bibnamefont {Marcus}}, \ and\ \bibinfo {author} {\bibfnamefont
  {F.}~\bibnamefont {Nichele}},\ }\href {\doibase
  10.1103/PhysRevLett.121.256803} {\bibfield  {journal} {\bibinfo  {journal}
  {Phys. Rev. Lett.}\ }\textbf {\bibinfo {volume} {121}},\ \bibinfo {pages}
  {256803} (\bibinfo {year} {2018})}\BibitemShut {NoStop}%
\bibitem [{\citenamefont {Chang}\ \emph {et~al.}(2013)\citenamefont {Chang},
  \citenamefont {Manucharyan}, \citenamefont {Jespersen}, \citenamefont
  {Nygård},\ and\ \citenamefont {Marcus}}]{chang_tunneling_2013}%
  \BibitemOpen
  \bibfield  {author} {\bibinfo {author} {\bibfnamefont {W.}~\bibnamefont
  {Chang}}, \bibinfo {author} {\bibfnamefont {V.~E.}\ \bibnamefont
  {Manucharyan}}, \bibinfo {author} {\bibfnamefont {T.~S.}\ \bibnamefont
  {Jespersen}}, \bibinfo {author} {\bibfnamefont {J.}~\bibnamefont {Nygård}},
  \ and\ \bibinfo {author} {\bibfnamefont {C.~M.}\ \bibnamefont {Marcus}},\
  }\href {\doibase 10.1103/PhysRevLett.110.217005} {\bibfield  {journal}
  {\bibinfo  {journal} {Physical Review Letters}\ }\textbf {\bibinfo {volume}
  {110}},\ \bibinfo {pages} {217005} (\bibinfo {year} {2013})}\BibitemShut
  {NoStop}%
\bibitem [{\citenamefont {Jellinggaard}\ \emph {et~al.}(2016)\citenamefont
  {Jellinggaard}, \citenamefont {Grove-Rasmussen}, \citenamefont {Madsen},\
  and\ \citenamefont {Nygård}}]{jellinggaard_ysr_2016}%
  \BibitemOpen
  \bibfield  {author} {\bibinfo {author} {\bibfnamefont {A.}~\bibnamefont
  {Jellinggaard}}, \bibinfo {author} {\bibfnamefont {K.}~\bibnamefont
  {Grove-Rasmussen}}, \bibinfo {author} {\bibfnamefont {M.~H.}\ \bibnamefont
  {Madsen}}, \ and\ \bibinfo {author} {\bibfnamefont {J.}~\bibnamefont
  {Nygård}},\ }\href {\doibase 10.1103/PhysRevB.94.064520} {\bibfield
  {journal} {\bibinfo  {journal} {Phys. Rev. B}\ }\textbf {\bibinfo {volume}
  {94}},\ \bibinfo {pages} {064520} (\bibinfo {year} {2016})}\BibitemShut
  {NoStop}%
\bibitem [{\citenamefont {Kürtössy}\ \emph {et~al.}(2021)\citenamefont
  {Kürtössy}, \citenamefont {Scherübl}, \citenamefont {Fülöp},
  \citenamefont {Lukács}, \citenamefont {Kanne}, \citenamefont {Nygård},
  \citenamefont {Makk},\ and\ \citenamefont {Csonka}}]{kurtossy_andreev_2021}%
  \BibitemOpen
  \bibfield  {author} {\bibinfo {author} {\bibfnamefont {O.}~\bibnamefont
  {Kürtössy}}, \bibinfo {author} {\bibfnamefont {Z.}~\bibnamefont
  {Scherübl}}, \bibinfo {author} {\bibfnamefont {G.}~\bibnamefont {Fülöp}},
  \bibinfo {author} {\bibfnamefont {I.~E.}\ \bibnamefont {Lukács}}, \bibinfo
  {author} {\bibfnamefont {T.}~\bibnamefont {Kanne}}, \bibinfo {author}
  {\bibfnamefont {J.}~\bibnamefont {Nygård}}, \bibinfo {author} {\bibfnamefont
  {P.}~\bibnamefont {Makk}}, \ and\ \bibinfo {author} {\bibfnamefont
  {S.}~\bibnamefont {Csonka}},\ }\href {\doibase 10.1021/acs.nanolett.1c01956}
  {\bibfield  {journal} {\bibinfo  {journal} {Nano Lett.}\ }\textbf {\bibinfo
  {volume} {21}},\ \bibinfo {pages} {7929} (\bibinfo {year}
  {2021})}\BibitemShut {NoStop}%
\bibitem [{\citenamefont {Clarke}(2017)}]{clarke_experimentally_2017}%
  \BibitemOpen
  \bibfield  {author} {\bibinfo {author} {\bibfnamefont {D.~J.}\ \bibnamefont
  {Clarke}},\ }\href {\doibase 10.1103/PhysRevB.96.201109} {\bibfield
  {journal} {\bibinfo  {journal} {Phys. Rev. B}\ }\textbf {\bibinfo {volume}
  {96}},\ \bibinfo {pages} {201109} (\bibinfo {year} {2017})}\BibitemShut
  {NoStop}%
\bibitem [{\citenamefont {Prada}\ \emph {et~al.}(2017)\citenamefont {Prada},
  \citenamefont {Aguado},\ and\ \citenamefont
  {San-Jose}}]{prada_measuring_2017}%
  \BibitemOpen
  \bibfield  {author} {\bibinfo {author} {\bibfnamefont {E.}~\bibnamefont
  {Prada}}, \bibinfo {author} {\bibfnamefont {R.}~\bibnamefont {Aguado}}, \
  and\ \bibinfo {author} {\bibfnamefont {P.}~\bibnamefont {San-Jose}},\ }\href
  {\doibase 10.1103/PhysRevB.96.085418} {\bibfield  {journal} {\bibinfo
  {journal} {Phys. Rev. B}\ }\textbf {\bibinfo {volume} {96}},\ \bibinfo
  {pages} {085418} (\bibinfo {year} {2017})}\BibitemShut {NoStop}%
\bibitem [{\citenamefont {Peñaranda}\ \emph {et~al.}(2018)\citenamefont
  {Peñaranda}, \citenamefont {Aguado}, \citenamefont {San-Jose},\ and\
  \citenamefont {Prada}}]{penaranda_quantifying_2018}%
  \BibitemOpen
  \bibfield  {author} {\bibinfo {author} {\bibfnamefont {F.}~\bibnamefont
  {Peñaranda}}, \bibinfo {author} {\bibfnamefont {R.}~\bibnamefont {Aguado}},
  \bibinfo {author} {\bibfnamefont {P.}~\bibnamefont {San-Jose}}, \ and\
  \bibinfo {author} {\bibfnamefont {E.}~\bibnamefont {Prada}},\ }\href
  {\doibase 10.1103/PhysRevB.98.235406} {\bibfield  {journal} {\bibinfo
  {journal} {Phys. Rev. B}\ }\textbf {\bibinfo {volume} {98}},\ \bibinfo
  {pages} {235406} (\bibinfo {year} {2018})}\BibitemShut {NoStop}%
\bibitem [{\citenamefont {Deng}\ \emph {et~al.}(2018)\citenamefont {Deng},
  \citenamefont {Vaitiekėnas}, \citenamefont {Prada}, \citenamefont
  {San-Jose}, \citenamefont {Nygård}, \citenamefont {Krogstrup}, \citenamefont
  {Aguado},\ and\ \citenamefont {Marcus}}]{deng_nonlocality_2018}%
  \BibitemOpen
  \bibfield  {author} {\bibinfo {author} {\bibfnamefont {M.-T.}\ \bibnamefont
  {Deng}}, \bibinfo {author} {\bibfnamefont {S.}~\bibnamefont {Vaitiekėnas}},
  \bibinfo {author} {\bibfnamefont {E.}~\bibnamefont {Prada}}, \bibinfo
  {author} {\bibfnamefont {P.}~\bibnamefont {San-Jose}}, \bibinfo {author}
  {\bibfnamefont {J.}~\bibnamefont {Nygård}}, \bibinfo {author} {\bibfnamefont
  {P.}~\bibnamefont {Krogstrup}}, \bibinfo {author} {\bibfnamefont
  {R.}~\bibnamefont {Aguado}}, \ and\ \bibinfo {author} {\bibfnamefont {C.~M.}\
  \bibnamefont {Marcus}},\ }\href {\doibase 10.1103/PhysRevB.98.085125}
  {\bibfield  {journal} {\bibinfo  {journal} {Phys. Rev. B}\ }\textbf {\bibinfo
  {volume} {98}},\ \bibinfo {pages} {085125} (\bibinfo {year}
  {2018})}\BibitemShut {NoStop}%
\bibitem [{\citenamefont {Pöschl}\ \emph
  {et~al.}(2022{\natexlab{a}})\citenamefont {Pöschl}, \citenamefont
  {Danilenko}, \citenamefont {Sabonis}, \citenamefont {Kristjuhan},
  \citenamefont {Lindemann}, \citenamefont {Thomas}, \citenamefont {Manfra},\
  and\ \citenamefont {Marcus}}]{PoschlPC2022}%
  \BibitemOpen
  \bibfield  {author} {\bibinfo {author} {\bibfnamefont {A.}~\bibnamefont
  {Pöschl}}, \bibinfo {author} {\bibfnamefont {A.}~\bibnamefont {Danilenko}},
  \bibinfo {author} {\bibfnamefont {D.}~\bibnamefont {Sabonis}}, \bibinfo
  {author} {\bibfnamefont {K.}~\bibnamefont {Kristjuhan}}, \bibinfo {author}
  {\bibfnamefont {T.}~\bibnamefont {Lindemann}}, \bibinfo {author}
  {\bibfnamefont {C.}~\bibnamefont {Thomas}}, \bibinfo {author} {\bibfnamefont
  {M.~J.}\ \bibnamefont {Manfra}}, \ and\ \bibinfo {author} {\bibfnamefont
  {C.~M.}\ \bibnamefont {Marcus}},\ }\href {\doibase
  10.1103/PhysRevB.106.L161301} {\bibfield  {journal} {\bibinfo  {journal}
  {Phys. Rev. B}\ }\textbf {\bibinfo {volume} {106}},\ \bibinfo {pages}
  {L161301} (\bibinfo {year} {2022}{\natexlab{a}})}\BibitemShut {NoStop}%
\bibitem [{\citenamefont {J{\"u}nger}\ \emph {et~al.}(2019)\citenamefont
  {J{\"u}nger}, \citenamefont {Baumgartner}, \citenamefont {Delagrange},
  \citenamefont {Chevallier}, \citenamefont {Lehmann}, \citenamefont {Nilsson},
  \citenamefont {Dick}, \citenamefont {Thelander},\ and\ \citenamefont
  {Sch{\"o}nenberger}}]{junger_dot_2019}%
  \BibitemOpen
  \bibfield  {author} {\bibinfo {author} {\bibfnamefont {C.}~\bibnamefont
  {J{\"u}nger}}, \bibinfo {author} {\bibfnamefont {A.}~\bibnamefont
  {Baumgartner}}, \bibinfo {author} {\bibfnamefont {R.}~\bibnamefont
  {Delagrange}}, \bibinfo {author} {\bibfnamefont {D.}~\bibnamefont
  {Chevallier}}, \bibinfo {author} {\bibfnamefont {S.}~\bibnamefont {Lehmann}},
  \bibinfo {author} {\bibfnamefont {M.}~\bibnamefont {Nilsson}}, \bibinfo
  {author} {\bibfnamefont {K.~A.}\ \bibnamefont {Dick}}, \bibinfo {author}
  {\bibfnamefont {C.}~\bibnamefont {Thelander}}, \ and\ \bibinfo {author}
  {\bibfnamefont {C.}~\bibnamefont {Sch{\"o}nenberger}},\ }\href {\doibase
  10.1038/s42005-019-0162-4} {\bibfield  {journal} {\bibinfo  {journal}
  {Commun. Phys.}\ }\textbf {\bibinfo {volume} {2}},\ \bibinfo {pages} {76}
  (\bibinfo {year} {2019})}\BibitemShut {NoStop}%
\bibitem [{\citenamefont {Thomas}\ \emph {et~al.}(2021)\citenamefont {Thomas},
  \citenamefont {Nilsson}, \citenamefont {Ciaccia}, \citenamefont {J\"unger},
  \citenamefont {Rossi}, \citenamefont {Zannier}, \citenamefont {Sorba},
  \citenamefont {Baumgartner},\ and\ \citenamefont
  {Sch\"onenberger}}]{thomas2021}%
  \BibitemOpen
  \bibfield  {author} {\bibinfo {author} {\bibfnamefont {F.~S.}\ \bibnamefont
  {Thomas}}, \bibinfo {author} {\bibfnamefont {M.}~\bibnamefont {Nilsson}},
  \bibinfo {author} {\bibfnamefont {C.}~\bibnamefont {Ciaccia}}, \bibinfo
  {author} {\bibfnamefont {C.}~\bibnamefont {J\"unger}}, \bibinfo {author}
  {\bibfnamefont {F.}~\bibnamefont {Rossi}}, \bibinfo {author} {\bibfnamefont
  {V.}~\bibnamefont {Zannier}}, \bibinfo {author} {\bibfnamefont
  {L.}~\bibnamefont {Sorba}}, \bibinfo {author} {\bibfnamefont
  {A.}~\bibnamefont {Baumgartner}}, \ and\ \bibinfo {author} {\bibfnamefont
  {C.}~\bibnamefont {Sch\"onenberger}},\ }\href {\doibase
  10.1103/PhysRevB.104.115415} {\bibfield  {journal} {\bibinfo  {journal}
  {Phys. Rev. B}\ }\textbf {\bibinfo {volume} {104}},\ \bibinfo {pages}
  {115415} (\bibinfo {year} {2021})}\BibitemShut {NoStop}%
\bibitem [{\citenamefont {Gramich}\ \emph {et~al.}(2016)\citenamefont
  {Gramich}, \citenamefont {Baumgartner},\ and\ \citenamefont
  {Schönenberger}}]{gramich_subgap_2016}%
  \BibitemOpen
  \bibfield  {author} {\bibinfo {author} {\bibfnamefont {J.}~\bibnamefont
  {Gramich}}, \bibinfo {author} {\bibfnamefont {A.}~\bibnamefont
  {Baumgartner}}, \ and\ \bibinfo {author} {\bibfnamefont {C.}~\bibnamefont
  {Schönenberger}},\ }\href {\doibase 10.1063/1.4948352} {\bibfield  {journal}
  {\bibinfo  {journal} {Appl. Phys. Lett.}\ }\textbf {\bibinfo {volume}
  {108}},\ \bibinfo {pages} {172604} (\bibinfo {year} {2016})}\BibitemShut
  {NoStop}%
\bibitem [{\citenamefont {Hanson}\ \emph {et~al.}(2003)\citenamefont {Hanson},
  \citenamefont {Witkamp}, \citenamefont {Vandersypen}, \citenamefont {van
  Beveren}, \citenamefont {Elzerman},\ and\ \citenamefont
  {Kouwenhoven}}]{hanson2003}%
  \BibitemOpen
  \bibfield  {author} {\bibinfo {author} {\bibfnamefont {R.}~\bibnamefont
  {Hanson}}, \bibinfo {author} {\bibfnamefont {B.}~\bibnamefont {Witkamp}},
  \bibinfo {author} {\bibfnamefont {L.~M.~K.}\ \bibnamefont {Vandersypen}},
  \bibinfo {author} {\bibfnamefont {L.~H.~W.}\ \bibnamefont {van Beveren}},
  \bibinfo {author} {\bibfnamefont {J.~M.}\ \bibnamefont {Elzerman}}, \ and\
  \bibinfo {author} {\bibfnamefont {L.~P.}\ \bibnamefont {Kouwenhoven}},\
  }\href {\doibase 10.1103/PhysRevLett.91.196802} {\bibfield  {journal}
  {\bibinfo  {journal} {Phys. Rev. Lett.}\ }\textbf {\bibinfo {volume} {91}},\
  \bibinfo {pages} {196802} (\bibinfo {year} {2003})}\BibitemShut {NoStop}%
\bibitem [{\citenamefont {Wang}\ \emph {et~al.}(2022)\citenamefont {Wang},
  \citenamefont {Dvir}, \citenamefont {van Loo}, \citenamefont {Mazur},
  \citenamefont {Gazibegovic}, \citenamefont {Badawy}, \citenamefont {Bakkers},
  \citenamefont {Kouwenhoven},\ and\ \citenamefont
  {de~Lange}}]{wang_nonlocal_2022}%
  \BibitemOpen
  \bibfield  {author} {\bibinfo {author} {\bibfnamefont {G.}~\bibnamefont
  {Wang}}, \bibinfo {author} {\bibfnamefont {T.}~\bibnamefont {Dvir}}, \bibinfo
  {author} {\bibfnamefont {N.}~\bibnamefont {van Loo}}, \bibinfo {author}
  {\bibfnamefont {G.~P.}\ \bibnamefont {Mazur}}, \bibinfo {author}
  {\bibfnamefont {S.}~\bibnamefont {Gazibegovic}}, \bibinfo {author}
  {\bibfnamefont {G.}~\bibnamefont {Badawy}}, \bibinfo {author} {\bibfnamefont
  {E.~P. A.~M.}\ \bibnamefont {Bakkers}}, \bibinfo {author} {\bibfnamefont
  {L.~P.}\ \bibnamefont {Kouwenhoven}}, \ and\ \bibinfo {author} {\bibfnamefont
  {G.}~\bibnamefont {de~Lange}},\ }\href {\doibase 10.1103/PhysRevB.106.064503}
  {\bibfield  {journal} {\bibinfo  {journal} {Physical Review B}\ }\textbf
  {\bibinfo {volume} {106}},\ \bibinfo {pages} {064503} (\bibinfo {year}
  {2022})}\BibitemShut {NoStop}%
\bibitem [{\citenamefont {van Driel}\ \emph {et~al.}(2022)\citenamefont {van
  Driel}, \citenamefont {Wang}, \citenamefont {Bordin}, \citenamefont {van
  Loo}, \citenamefont {Zatelli}, \citenamefont {Mazur}, \citenamefont {Xu},
  \citenamefont {Gazibegovic}, \citenamefont {Badawi}, \citenamefont {Bakkers},
  \citenamefont {Kouwenhoven},\ and\ \citenamefont {Dvir}}]{vanDriel}%
  \BibitemOpen
  \bibfield  {author} {\bibinfo {author} {\bibfnamefont {D.}~\bibnamefont {van
  Driel}}, \bibinfo {author} {\bibfnamefont {G.}~\bibnamefont {Wang}}, \bibinfo
  {author} {\bibfnamefont {A.}~\bibnamefont {Bordin}}, \bibinfo {author}
  {\bibfnamefont {N.}~\bibnamefont {van Loo}}, \bibinfo {author} {\bibfnamefont
  {F.}~\bibnamefont {Zatelli}}, \bibinfo {author} {\bibfnamefont {G.~P.}\
  \bibnamefont {Mazur}}, \bibinfo {author} {\bibfnamefont {D.}~\bibnamefont
  {Xu}}, \bibinfo {author} {\bibfnamefont {S.}~\bibnamefont {Gazibegovic}},
  \bibinfo {author} {\bibfnamefont {G.}~\bibnamefont {Badawi}}, \bibinfo
  {author} {\bibfnamefont {E.~P. A.~M.}\ \bibnamefont {Bakkers}}, \bibinfo
  {author} {\bibfnamefont {L.~P.}\ \bibnamefont {Kouwenhoven}}, \ and\ \bibinfo
  {author} {\bibfnamefont {T.}~\bibnamefont {Dvir}},\ }\href
  {https://arxiv.org/abs/2212.10241} {\bibfield  {journal} {\bibinfo  {journal}
  {arXiv:2212.10241}\ } (\bibinfo {year} {2022})}\BibitemShut {NoStop}%
\bibitem [{\citenamefont {Meservey}\ and\ \citenamefont
  {Tedrow}(1994)}]{meservey1994}%
  \BibitemOpen
  \bibfield  {author} {\bibinfo {author} {\bibfnamefont {R.}~\bibnamefont
  {Meservey}}\ and\ \bibinfo {author} {\bibfnamefont {P.~M.}\ \bibnamefont
  {Tedrow}},\ }\href {\doibase 10.1016/0370-1573(94)90105-8} {\bibfield
  {journal} {\bibinfo  {journal} {Physics Reports}\ }\textbf {\bibinfo {volume}
  {238}},\ \bibinfo {pages} {173} (\bibinfo {year} {1994})}\BibitemShut
  {NoStop}%
\bibitem [{\citenamefont {Gramich}\ \emph {et~al.}(2017)\citenamefont
  {Gramich}, \citenamefont {Baumgartner},\ and\ \citenamefont
  {Schönenberger}}]{gramich_andreev_2017}%
  \BibitemOpen
  \bibfield  {author} {\bibinfo {author} {\bibfnamefont {J.}~\bibnamefont
  {Gramich}}, \bibinfo {author} {\bibfnamefont {A.}~\bibnamefont
  {Baumgartner}}, \ and\ \bibinfo {author} {\bibfnamefont {C.}~\bibnamefont
  {Schönenberger}},\ }\href {\doibase 10.1103/PhysRevB.96.195418} {\bibfield
  {journal} {\bibinfo  {journal} {Physical Review B}\ }\textbf {\bibinfo
  {volume} {96}},\ \bibinfo {pages} {195418} (\bibinfo {year}
  {2017})}\BibitemShut {NoStop}%
\bibitem [{\citenamefont {Recher}\ \emph {et~al.}(2000)\citenamefont {Recher},
  \citenamefont {Sukhorukov},\ and\ \citenamefont {Loss}}]{recher2000}%
  \BibitemOpen
  \bibfield  {author} {\bibinfo {author} {\bibfnamefont {P.}~\bibnamefont
  {Recher}}, \bibinfo {author} {\bibfnamefont {E.~V.}\ \bibnamefont
  {Sukhorukov}}, \ and\ \bibinfo {author} {\bibfnamefont {D.}~\bibnamefont
  {Loss}},\ }\href {\doibase 10.1103/PhysRevLett.85.1962} {\bibfield  {journal}
  {\bibinfo  {journal} {Phys. Rev. Lett.}\ }\textbf {\bibinfo {volume} {85}},\
  \bibinfo {pages} {1962} (\bibinfo {year} {2000})}\BibitemShut {NoStop}%
\bibitem [{\citenamefont {Antipov}\ \emph {et~al.}(2018)\citenamefont
  {Antipov}, \citenamefont {Bargerbos}, \citenamefont {Winkler}, \citenamefont
  {Bauer}, \citenamefont {Rossi},\ and\ \citenamefont
  {Lutchyn}}]{antipov_gfactor_2018}%
  \BibitemOpen
  \bibfield  {author} {\bibinfo {author} {\bibfnamefont {A.~E.}\ \bibnamefont
  {Antipov}}, \bibinfo {author} {\bibfnamefont {A.}~\bibnamefont {Bargerbos}},
  \bibinfo {author} {\bibfnamefont {G.~W.}\ \bibnamefont {Winkler}}, \bibinfo
  {author} {\bibfnamefont {B.}~\bibnamefont {Bauer}}, \bibinfo {author}
  {\bibfnamefont {E.}~\bibnamefont {Rossi}}, \ and\ \bibinfo {author}
  {\bibfnamefont {R.~M.}\ \bibnamefont {Lutchyn}},\ }\href {\doibase
  10.1103/PhysRevX.8.031041} {\bibfield  {journal} {\bibinfo  {journal} {Phys.
  Rev. X}\ }\textbf {\bibinfo {volume} {8}},\ \bibinfo {pages} {031041}
  (\bibinfo {year} {2018})}\BibitemShut {NoStop}%
\bibitem [{\citenamefont {Smith~III}\ and\ \citenamefont
  {Fang}(1987)}]{smith_iii_g_1987}%
  \BibitemOpen
  \bibfield  {author} {\bibinfo {author} {\bibfnamefont {T.~P.}\ \bibnamefont
  {Smith~III}}\ and\ \bibinfo {author} {\bibfnamefont {F.~F.}\ \bibnamefont
  {Fang}},\ }\href {\doibase 10.1103/PhysRevB.35.7729} {\bibfield  {journal}
  {\bibinfo  {journal} {Physical Review B}\ }\textbf {\bibinfo {volume} {35}},\
  \bibinfo {pages} {7729} (\bibinfo {year} {1987})}\BibitemShut {NoStop}%
\bibitem [{\citenamefont {Soulen}\ \emph {et~al.}(1998)\citenamefont {Soulen},
  \citenamefont {Byers}, \citenamefont {Osofsky}, \citenamefont {Nadgorny},
  \citenamefont {Ambrose}, \citenamefont {Cheng}, \citenamefont {Broussard},
  \citenamefont {Tanaka}, \citenamefont {Nowak}, \citenamefont {Moodera},
  \citenamefont {Barry},\ and\ \citenamefont {Coey}}]{soulen_measuring_1998}%
  \BibitemOpen
  \bibfield  {author} {\bibinfo {author} {\bibfnamefont {R.~J.}\ \bibnamefont
  {Soulen}}, \bibinfo {author} {\bibfnamefont {J.~M.}\ \bibnamefont {Byers}},
  \bibinfo {author} {\bibfnamefont {M.~S.}\ \bibnamefont {Osofsky}}, \bibinfo
  {author} {\bibfnamefont {B.}~\bibnamefont {Nadgorny}}, \bibinfo {author}
  {\bibfnamefont {T.}~\bibnamefont {Ambrose}}, \bibinfo {author} {\bibfnamefont
  {S.~F.}\ \bibnamefont {Cheng}}, \bibinfo {author} {\bibfnamefont {P.~R.}\
  \bibnamefont {Broussard}}, \bibinfo {author} {\bibfnamefont {C.~T.}\
  \bibnamefont {Tanaka}}, \bibinfo {author} {\bibfnamefont {J.}~\bibnamefont
  {Nowak}}, \bibinfo {author} {\bibfnamefont {J.~S.}\ \bibnamefont {Moodera}},
  \bibinfo {author} {\bibfnamefont {A.}~\bibnamefont {Barry}}, \ and\ \bibinfo
  {author} {\bibfnamefont {J.~M.~D.}\ \bibnamefont {Coey}},\ }\href {\doibase
  10.1126/science.282.5386.85} {\bibfield  {journal} {\bibinfo  {journal}
  {Science}\ }\textbf {\bibinfo {volume} {282}},\ \bibinfo {pages} {85}
  (\bibinfo {year} {1998})}\BibitemShut {NoStop}%
\bibitem [{\citenamefont {Lee}\ \emph {et~al.}(2014)\citenamefont {Lee},
  \citenamefont {Jiang}, \citenamefont {Houzet}, \citenamefont {Aguado},
  \citenamefont {Lieber},\ and\ \citenamefont
  {de~Franceschi}}]{lee_spin-resolved_2014}%
  \BibitemOpen
  \bibfield  {author} {\bibinfo {author} {\bibfnamefont {E.~J.~H.}\
  \bibnamefont {Lee}}, \bibinfo {author} {\bibfnamefont {X.}~\bibnamefont
  {Jiang}}, \bibinfo {author} {\bibfnamefont {M.}~\bibnamefont {Houzet}},
  \bibinfo {author} {\bibfnamefont {R.}~\bibnamefont {Aguado}}, \bibinfo
  {author} {\bibfnamefont {C.~M.}\ \bibnamefont {Lieber}}, \ and\ \bibinfo
  {author} {\bibfnamefont {S.}~\bibnamefont {de~Franceschi}},\ }\href@noop {}
  {\bibfield  {journal} {\bibinfo  {journal} {Nat.~Nano.}\ }\textbf {\bibinfo
  {volume} {9 1}},\ \bibinfo {pages} {79} (\bibinfo {year} {2014})}\BibitemShut
  {NoStop}%
\bibitem [{\citenamefont {Pillet}\ \emph {et~al.}(2013)\citenamefont {Pillet},
  \citenamefont {Joyez}, \citenamefont {Žitko},\ and\ \citenamefont
  {Goffman}}]{pillet_tunneling_2013}%
  \BibitemOpen
  \bibfield  {author} {\bibinfo {author} {\bibfnamefont {J.-D.}\ \bibnamefont
  {Pillet}}, \bibinfo {author} {\bibfnamefont {P.}~\bibnamefont {Joyez}},
  \bibinfo {author} {\bibfnamefont {R.}~\bibnamefont {Žitko}}, \ and\ \bibinfo
  {author} {\bibfnamefont {M.~F.}\ \bibnamefont {Goffman}},\ }\href {\doibase
  10.1103/PhysRevB.88.045101} {\bibfield  {journal} {\bibinfo  {journal}
  {Physical Review B}\ }\textbf {\bibinfo {volume} {88}},\ \bibinfo {pages}
  {045101} (\bibinfo {year} {2013})}\BibitemShut {NoStop}%
\bibitem [{\citenamefont {Schindele}\ \emph {et~al.}(2014)\citenamefont
  {Schindele}, \citenamefont {Baumgartner}, \citenamefont {Maurand},
  \citenamefont {Weiss},\ and\ \citenamefont
  {Schönenberger}}]{schindele_nonlocal_2014}%
  \BibitemOpen
  \bibfield  {author} {\bibinfo {author} {\bibfnamefont {J.}~\bibnamefont
  {Schindele}}, \bibinfo {author} {\bibfnamefont {A.}~\bibnamefont
  {Baumgartner}}, \bibinfo {author} {\bibfnamefont {R.}~\bibnamefont
  {Maurand}}, \bibinfo {author} {\bibfnamefont {M.}~\bibnamefont {Weiss}}, \
  and\ \bibinfo {author} {\bibfnamefont {C.}~\bibnamefont {Schönenberger}},\
  }\href {\doibase 10.1103/PhysRevB.89.045422} {\bibfield  {journal} {\bibinfo
  {journal} {Physical Review B}\ }\textbf {\bibinfo {volume} {89}},\ \bibinfo
  {pages} {045422} (\bibinfo {year} {2014})}\BibitemShut {NoStop}%
\bibitem [{\citenamefont {Pöschl}\ \emph
  {et~al.}(2022{\natexlab{b}})\citenamefont {Pöschl}, \citenamefont
  {Danilenko}, \citenamefont {Sabonis}, \citenamefont {Kristjuhan},
  \citenamefont {Lindemann}, \citenamefont {Thomas}, \citenamefont {Manfra},\
  and\ \citenamefont {Marcus}}]{poschl_nonlocal_2022}%
  \BibitemOpen
  \bibfield  {author} {\bibinfo {author} {\bibfnamefont {A.}~\bibnamefont
  {Pöschl}}, \bibinfo {author} {\bibfnamefont {A.}~\bibnamefont {Danilenko}},
  \bibinfo {author} {\bibfnamefont {D.}~\bibnamefont {Sabonis}}, \bibinfo
  {author} {\bibfnamefont {K.}~\bibnamefont {Kristjuhan}}, \bibinfo {author}
  {\bibfnamefont {T.}~\bibnamefont {Lindemann}}, \bibinfo {author}
  {\bibfnamefont {C.}~\bibnamefont {Thomas}}, \bibinfo {author} {\bibfnamefont
  {M.~J.}\ \bibnamefont {Manfra}}, \ and\ \bibinfo {author} {\bibfnamefont
  {C.~M.}\ \bibnamefont {Marcus}},\ }\href {\doibase
  10.1103/PhysRevB.106.L241301} {\bibfield  {journal} {\bibinfo  {journal}
  {Physical Review B}\ }\textbf {\bibinfo {volume} {106}},\ \bibinfo {pages}
  {L241301} (\bibinfo {year} {2022}{\natexlab{b}})}\BibitemShut {NoStop}%
\bibitem [{\citenamefont {Danon}\ \emph {et~al.}(2020)\citenamefont {Danon},
  \citenamefont {Hellenes}, \citenamefont {Hansen}, \citenamefont {Casparis},
  \citenamefont {Higginbotham},\ and\ \citenamefont
  {Flensberg}}]{karsten_nl_spectroscopy}%
  \BibitemOpen
  \bibfield  {author} {\bibinfo {author} {\bibfnamefont {J.}~\bibnamefont
  {Danon}}, \bibinfo {author} {\bibfnamefont {A.~B.}\ \bibnamefont {Hellenes}},
  \bibinfo {author} {\bibfnamefont {E.~B.}\ \bibnamefont {Hansen}}, \bibinfo
  {author} {\bibfnamefont {L.}~\bibnamefont {Casparis}}, \bibinfo {author}
  {\bibfnamefont {A.~P.}\ \bibnamefont {Higginbotham}}, \ and\ \bibinfo
  {author} {\bibfnamefont {K.}~\bibnamefont {Flensberg}},\ }\href {\doibase
  10.1103/PhysRevLett.124.036801} {\bibfield  {journal} {\bibinfo  {journal}
  {Phys. Rev. Lett.}\ }\textbf {\bibinfo {volume} {124}},\ \bibinfo {pages}
  {036801} (\bibinfo {year} {2020})}\BibitemShut {NoStop}%
\bibitem [{\citenamefont {Puglia}\ \emph {et~al.}(2021)\citenamefont {Puglia},
  \citenamefont {Martinez}, \citenamefont {Ménard}, \citenamefont {Pöschl},
  \citenamefont {Gronin}, \citenamefont {Gardner}, \citenamefont {Kallaher},
  \citenamefont {Manfra}, \citenamefont {Marcus}, \citenamefont
  {Higginbotham},\ and\ \citenamefont {Casparis}}]{denise_nl_gapclosing}%
  \BibitemOpen
  \bibfield  {author} {\bibinfo {author} {\bibfnamefont {D.}~\bibnamefont
  {Puglia}}, \bibinfo {author} {\bibfnamefont {E.~A.}\ \bibnamefont
  {Martinez}}, \bibinfo {author} {\bibfnamefont {G.~C.}\ \bibnamefont
  {Ménard}}, \bibinfo {author} {\bibfnamefont {A.}~\bibnamefont {Pöschl}},
  \bibinfo {author} {\bibfnamefont {S.}~\bibnamefont {Gronin}}, \bibinfo
  {author} {\bibfnamefont {G.~C.}\ \bibnamefont {Gardner}}, \bibinfo {author}
  {\bibfnamefont {R.}~\bibnamefont {Kallaher}}, \bibinfo {author}
  {\bibfnamefont {M.~J.}\ \bibnamefont {Manfra}}, \bibinfo {author}
  {\bibfnamefont {C.~M.}\ \bibnamefont {Marcus}}, \bibinfo {author}
  {\bibfnamefont {A.~P.}\ \bibnamefont {Higginbotham}}, \ and\ \bibinfo
  {author} {\bibfnamefont {L.}~\bibnamefont {Casparis}},\ }\href {\doibase
  10.1103/PhysRevB.103.235201} {\bibfield  {journal} {\bibinfo  {journal}
  {Phys. Rev. B}\ }\textbf {\bibinfo {volume} {103}},\ \bibinfo {pages}
  {235201} (\bibinfo {year} {2021})}\BibitemShut {NoStop}%
\bibitem [{\citenamefont {Hanson}\ \emph {et~al.}(2004)\citenamefont {Hanson},
  \citenamefont {Vandersypen}, \citenamefont {van Beveren}, \citenamefont
  {Elzerman}, \citenamefont {Vink},\ and\ \citenamefont
  {Kouwenhoven}}]{hanson_bipolar_2004}%
  \BibitemOpen
  \bibfield  {author} {\bibinfo {author} {\bibfnamefont {R.}~\bibnamefont
  {Hanson}}, \bibinfo {author} {\bibfnamefont {L.~M.~K.}\ \bibnamefont
  {Vandersypen}}, \bibinfo {author} {\bibfnamefont {L.~H.~W.}\ \bibnamefont
  {van Beveren}}, \bibinfo {author} {\bibfnamefont {J.~M.}\ \bibnamefont
  {Elzerman}}, \bibinfo {author} {\bibfnamefont {I.~T.}\ \bibnamefont {Vink}},
  \ and\ \bibinfo {author} {\bibfnamefont {L.~P.}\ \bibnamefont
  {Kouwenhoven}},\ }\href {\doibase 10.1103/PhysRevB.70.241304} {\bibfield
  {journal} {\bibinfo  {journal} {Phys. Rev. B}\ }\textbf {\bibinfo {volume}
  {70}},\ \bibinfo {pages} {241304} (\bibinfo {year} {2004})}\BibitemShut
  {NoStop}%
\bibitem [{\citenamefont {Hanson}\ \emph {et~al.}(2007)\citenamefont {Hanson},
  \citenamefont {Kouwenhoven}, \citenamefont {Petta}, \citenamefont {Tarucha},\
  and\ \citenamefont {Vandersypen}}]{hanson_spins_2007}%
  \BibitemOpen
  \bibfield  {author} {\bibinfo {author} {\bibfnamefont {R.}~\bibnamefont
  {Hanson}}, \bibinfo {author} {\bibfnamefont {L.~P.}\ \bibnamefont
  {Kouwenhoven}}, \bibinfo {author} {\bibfnamefont {J.~R.}\ \bibnamefont
  {Petta}}, \bibinfo {author} {\bibfnamefont {S.}~\bibnamefont {Tarucha}}, \
  and\ \bibinfo {author} {\bibfnamefont {L.~M.~K.}\ \bibnamefont
  {Vandersypen}},\ }\href {\doibase 10.1103/RevModPhys.79.1217} {\bibfield
  {journal} {\bibinfo  {journal} {Rev. Mod. Phys.}\ }\textbf {\bibinfo {volume}
  {79}},\ \bibinfo {pages} {1217} (\bibinfo {year} {2007})}\BibitemShut
  {NoStop}%
\bibitem [{\citenamefont {Szumniak}\ \emph {et~al.}(2017)\citenamefont
  {Szumniak}, \citenamefont {Chevallier}, \citenamefont {Loss},\ and\
  \citenamefont {Klinovaja}}]{szumniak_spin_2017}%
  \BibitemOpen
  \bibfield  {author} {\bibinfo {author} {\bibfnamefont {P.}~\bibnamefont
  {Szumniak}}, \bibinfo {author} {\bibfnamefont {D.}~\bibnamefont
  {Chevallier}}, \bibinfo {author} {\bibfnamefont {D.}~\bibnamefont {Loss}}, \
  and\ \bibinfo {author} {\bibfnamefont {J.}~\bibnamefont {Klinovaja}},\ }\href
  {\doibase 10.1103/PhysRevB.96.041401} {\bibfield  {journal} {\bibinfo
  {journal} {Phys. Rev. B}\ }\textbf {\bibinfo {volume} {96}},\ \bibinfo
  {pages} {041401} (\bibinfo {year} {2017})}\BibitemShut {NoStop}%
\bibitem [{\citenamefont {Chevallier}\ \emph {et~al.}(2018)\citenamefont
  {Chevallier}, \citenamefont {Szumniak}, \citenamefont {Hoffman},
  \citenamefont {Loss},\ and\ \citenamefont
  {Klinovaja}}]{chevallier_bulkband_2018}%
  \BibitemOpen
  \bibfield  {author} {\bibinfo {author} {\bibfnamefont {D.}~\bibnamefont
  {Chevallier}}, \bibinfo {author} {\bibfnamefont {P.}~\bibnamefont
  {Szumniak}}, \bibinfo {author} {\bibfnamefont {S.}~\bibnamefont {Hoffman}},
  \bibinfo {author} {\bibfnamefont {D.}~\bibnamefont {Loss}}, \ and\ \bibinfo
  {author} {\bibfnamefont {J.}~\bibnamefont {Klinovaja}},\ }\href {\doibase
  10.1103/PhysRevB.97.045404} {\bibfield  {journal} {\bibinfo  {journal} {Phys.
  Rev. B}\ }\textbf {\bibinfo {volume} {97}},\ \bibinfo {pages} {045404}
  (\bibinfo {year} {2018})}\BibitemShut {NoStop}%
\end{thebibliography}%

%%%%%%%%%% Merge with supplemental materials %%%%%%%%%%
%\widetext
%\clearpage
%\begin{center}
%\textbf{\large Supplemental Materials: Title for main text}
%\end{center}
%%%%%%%%%% Merge with supplemenental materials %%%%%%%%%%
%%%%%%%%%% Prefix a "S" to all equations, figures, tables and reset the counter %%%%%%%%%%
%\setcounter{equation}{0}
%\setcounter{figure}{0}
%\setcounter{table}{0}
%\setcounter{page}{1}
%\makeatletter
%\renewcommand{\theequation}{S\arabic{equation}}
%\renewcommand{\thefigure}{S\arabic{figure}}
%\renewcommand{\bibnumfmt}[1]{[S#1]}
%\renewcommand{\citenumfont}[1]{S#1}
%%%%%%%%%% Prefix a "S" to all equations, figures, tables and reset the counter %%%%%%%%%%

%\section{Section 1}

\end{document}